\documentclass[a4paper,11pt]{article}

\usepackage[a4paper,left=2.73cm,right=2.7cm,top=3cm,bottom=3.5cm]{geometry}

\usepackage{graphicx}
\usepackage{epstopdf}
\usepackage{amsmath}
\usepackage{amsfonts}
\usepackage{amssymb}
\usepackage{color}
\usepackage{mathrsfs}
\usepackage[utf8]{inputenc}

\usepackage{slashed}

\usepackage{csquotes}

\usepackage{caption}

\usepackage{cite}

\usepackage{booktabs}

\usepackage[colorlinks=true,bookmarks,citecolor=blue,linkcolor=blue]{hyperref}

\usepackage{authblk}

\newcommand{\Ehigh}{\mathcal{E}_{\text{high}}}
\newcommand{\Elow}{\mathcal{E}_{\text{low}}}
\newcommand{\fig}[1]{Fig.~\ref{#1}}

\def\E{{\mathcal{E}}}
\def\Ezero{{\mathcal{E}\left(t=0\right)}}

\author{Maximilian Attems}
\affil{Theoretical Physics Department, CERN, CH-1211 Genève 23, Switzerland.}
\affil{Instituto Galego de F\'\i sica de Altas Enerx\'\i as (IGFAE), Universidade de Santiago de Compostela, 15782 Galicia, Spain.}
\begin{document}

\begin{titlepage}

\thispagestyle{empty}

\begin{flushright}
CERN-TH-2020-223
\end{flushright}

\vspace{40pt}
\begin{center}

\title{Holographic approach of the spinodal instability to criticality}
{\Large \textbf{\mbox{Holographic approach of the spinodal instability to criticality}}}
        \vspace{30pt}

{\large \bf  Maximilian Attems}

\vspace{25pt}

{\normalsize Theoretical Physics Department, CERN, CH-1211 Genève 23, Switzerland.}

\vspace{5pt}

{\normalsize Instituto Galego de F\'\i sica de Altas Enerx\'\i as (IGFAE), Universidade de Santiago de Compostela, 15782 Galicia, Spain.} \\

\vspace{45pt}

\abstract{
A smoking gun signature for a first-order phase transition with negative
speed of sound squared $c_s^2$ is the occurrence of a spinodal instability.
In the gauge/gravity duality it corresponds to a
Gregory-Laflamme type instability, which can be numerically simulated as the
evolution of unstable planar black branes.
Making use of holography its dynamics is studied far from and near a critical
point with the following results.
Near a critical point the interface between cold and hot stable phases, given
by its width and surface tension, is found to feature a wider phase separation
and a smaller surface tension.
Far away from a critical point the formation time of the spinodal instability
is reduced.
Across softer and harder phase transitions, it is demonstrated that mergers of
equilibrated peaks and unstable plateaux lead to the preferred final single
phase separated solution.
Finally, a new atypical setup with dissipation of a peak into a plateau is discovered.
In order to distinguish the inhomogeneous states I propose a new criterium
based on the maximum of the transverse pressure at the interface
which encodes phase-mixed peaks versus fully phase separated plateaux.
}

\end{center}
\end{titlepage}
\newpage
\tableofcontents
\section{Introduction}
Analytical holographic studies pioneered the modeling of heavy-ion collisions
with the emergence of collectivity~\cite{Janik:2005zt,Albacete:2008vs,Grumiller:2008va,Gubser:2008pc}.
Fast hydrodynamization, implying the early applicability of hydrodynamics to a relaxing fluid, got first established numerically in dynamical out-of-equilibrium holographic shock wave collisions~\cite{Chesler:2010bi,Casalderrey-Solana:2013aba,Casalderrey-Solana:2013sxa,Chesler:2015wra,Chesler:2015bba} and further corroborated in the presence of bulk viscosity~\cite{Janik:2015waa,Buchel:2015ofa,Buchel:2015saa,Rougemont:2015wca,Attems:2016ugt,Attems:2016tby,Czajka:2018bod,Czajka:2018egm,Attems:2017zam} to model the almost perfect fluid.
As the Relativistic Heavy Ion Collider is accumulating plenty of
experimental data in the ongoing beam energy scan~\cite{Bzdak:2019pkr} and the
upcoming experimental
Facility for Antiproton and Ion Research~\cite{Friese:2006dj,Durante:2019hzd}
will search the quantum chromodynamics phase diagram for the presumed critical
point,
theoretical studies on criticality garner new attention~\cite{Stephanov:1998dy,Stephanov:1999zu,Erlich:2005qh,DeWolfe:2010he,Athanasiou:2010kw,Alba:2017hhe,Critelli:2017oub,Brewer:2018abr,Rougemont:2018ivt,Critelli:2018osu,Du:2020bxp,Li:2020hau,Hoyos:2020hmq,Dore:2020jye,Mroczek:2020rpm,Nahrgang:2020yxm,Dexheimer:2020zzs}.
In the cool down of the formed quark-gluon plasma in hadronic collisions there is presumably a wide temperature range with large
baryon chemical potential to hit the first-order phase transition line, whose endpoint is the critical point.
A prime signal for such a first-order type phase transition is the spinodal instability.
The gauge/gravity duality~\cite{Maldacena:1997re} opens up the possibility to study the dynamics of such a phase transition, as the holographic dual~\cite{Buchel:2005nt} of the spinodal instability is the Gregory-Laflamme instability~\cite{Gregory:1993vy,Emparan:2001wn,Emparan:2006mm,Emparan:2009cs,Emparan:2009at,Figueras:2015hkb}, which is amenable to general relativity calculations.

In \cite{Attems:2017ezz} we discovered  a metastable inhomogeneous unstable solution and subsequently \cite{Janik:2017ykj} found the phase separation of the spinodal instability.
The spinodal instability develops in four stages: 1) exponential growth of the instability; 2) the reshaping; 3) the merger; 4) the preferred final solution~\cite{Attems:2019yqn,Bellantuono:2019wbn}.
During the evolution of the spinodal instability one encounters different types of inhomogeneous states.
With a field redefinition all stages of a strong spinodal instability and the
hydrodynamization of shockwave collisions near a critical point were
demonstrated to be described by
hydrodynamics~\cite{Attems:2018gou,Attems:2019yqn}.
A recent analysis discusses the finite size effects~\cite{Bea:2020ees} of the
periodical longitudinal direction on the stability or instability of the plasma.
A quite different setup involving plasma balls studies effects of the confined
phase~\cite{Bantilan:2020pay}, but there is no spinodal region.

The purpose of this paper is to vary criticality in order to see the dynamical
effects of different first-order phase transitions.
For this endeavour a Gregory-Laflamme type instability is evolved on the gravity
side to induce the spinodal instability on the gauge theory side.
Strictly speaking, below the critical temperature the Gregory-Laflamme
instability has only unstable regions.
In the holographic construction a Gregory-Laflamme \emph{type} instability is
induced by the non-trivial potential of a weakly-coupled scalar field and is
unstable only in a certain temperature range.
To simplify the construction I consider a setup with no conserved charges,
hence the critical point or the phase transition will lie on the temperature
axes.
Another distinction to quantum-chromodynamics besides the class of the phase
transitions considered is the deconfined nature of the cold stable phase.
The strength of the holographic approach is to be able to extract and simulate
universal features of a two phase system in the limit of strong coupling.
The focus of this paper is on the dynamical real-time features of the spinodal instability while approaching a critical point.
Section 2 discusses the dual setup, in particular how to simulate on the
gravity side the Gregory-Laflamme type instability:
One evolves the Einstein equations in the bulk and reads off the boundary data
describing the spinodal instability on the gauge theory side.
If the first order phase transition occures far from (near) the critical point
one speaks of a strong (soft) first order phase transition.
Section 3 introduces a new criterium which is used to distinguish  the
inhomogeneous states - plateaux, peaks, valleys and gorges - in the relevant
stages of the spinodal instability (reshaping, merger and final).
This allows us in section 4 to characterize for different criticality
the shape and surface tension of the interface seen in the preferred and
settled final states of the spinodal instability.
In section 5  several dynamical real-time setups of the
spinodal instability with varying criticality are demonstrated,
including a newly revealed dissipative process of a peak into a plateaux.
Finally, I discuss the implications of the varying
criticality for the spinodal instability in section 6.

\section{Setup}
The non-conformal bottom-up model \cite{Attems:2016ugt,Attems:2017ezz} employed here, is described by Einstein Hilbert action coupled to a scalar with a non trivial potential $V(\phi)$
\begin{align}
S=\frac{2}{\kappa_{5}^{2}} \int d^{5} x \sqrt{-g} \left[ \frac{1}{4} \mathcal{R} - \frac{1}{2} \left( \nabla \phi \right) ^2 - V(\phi) \right] \,,
\end{align}
with $\kappa_5$ the five-dimensional Newton constant.
The chosen potential $V(\phi)$ is conformal in the ultra-violet, where the spacetime is Anti-de-Sitter (AdS), and has a minimum corresponding to an infrared fixed point in the gauge theory.
The potential is derived from a superpotential~\cite{Bianchi:2001kw}:
\begin{align}
\label{eq:superpotential}
	L W(\phi)=-\frac{3}{2}-\frac{\phi^2}{2}-\frac{\phi^4}{4\phi_M^2} \,,
\end{align}
where $L$ is the radius of the AdS solution in the ultra-violet.
The first two terms in \eqref{eq:superpotential} are fixed by
AdS and the dimension of the \enquote{dual} scalar operator.
The third term is responsible for the non-conformality and the appearance
of a critical point.
The potential has a single parameter $\phi_{\rm M}$, whose value defines the transition type.
For a subcritical value of $\phi_{\rm M}$ the transition is a first order phase transition, which then turns to a critical point, and finally for a supercritical value to a cross-over phase transition.
This yields the following scalar potential
\begin{align}
\label{eq:potential}
L^2 V(\phi) = -3 -\frac{3}{2}\phi^2 - \frac{1}{3} \phi^4 + \left(\frac{1}{2\phi_{\rm M}^4} {- \frac{1}{3\phi_{\rm M}^2}}\right) \phi^6- \frac{1}{12\phi_{\rm M}^4} \phi^8 \,,
\end{align}
where the term $-\frac{1}{3\phi_{\rm M}^2} \phi^6$ is responsible for critical behaviour.\\
In what follows, the numerical procedure outlined in \cite{Chesler:2013lia,vanderSchee:2014qwa,Attems:2017zam} is used and
I will set the dimensionless parameter of the potential to the subcritical values\footnote{
Supercritical values of $\phi_{\rm M}$ lead to a cross-over phase transition with no spinodal instability, see~\cite{Attems:2017zam} for the relaxation channels and out-of-equilibrium properties.
}
\begin{align}
\label{eq:phimvalues}
\phi_{\rm M} = \left\{2.25, 2.3, 2.35, 2.4, 2.45\right\}.
\end{align}
\begin{figure*}[t]
\begin{center}
\begin{tabular}{cc}
\includegraphics[width=.49\textwidth]{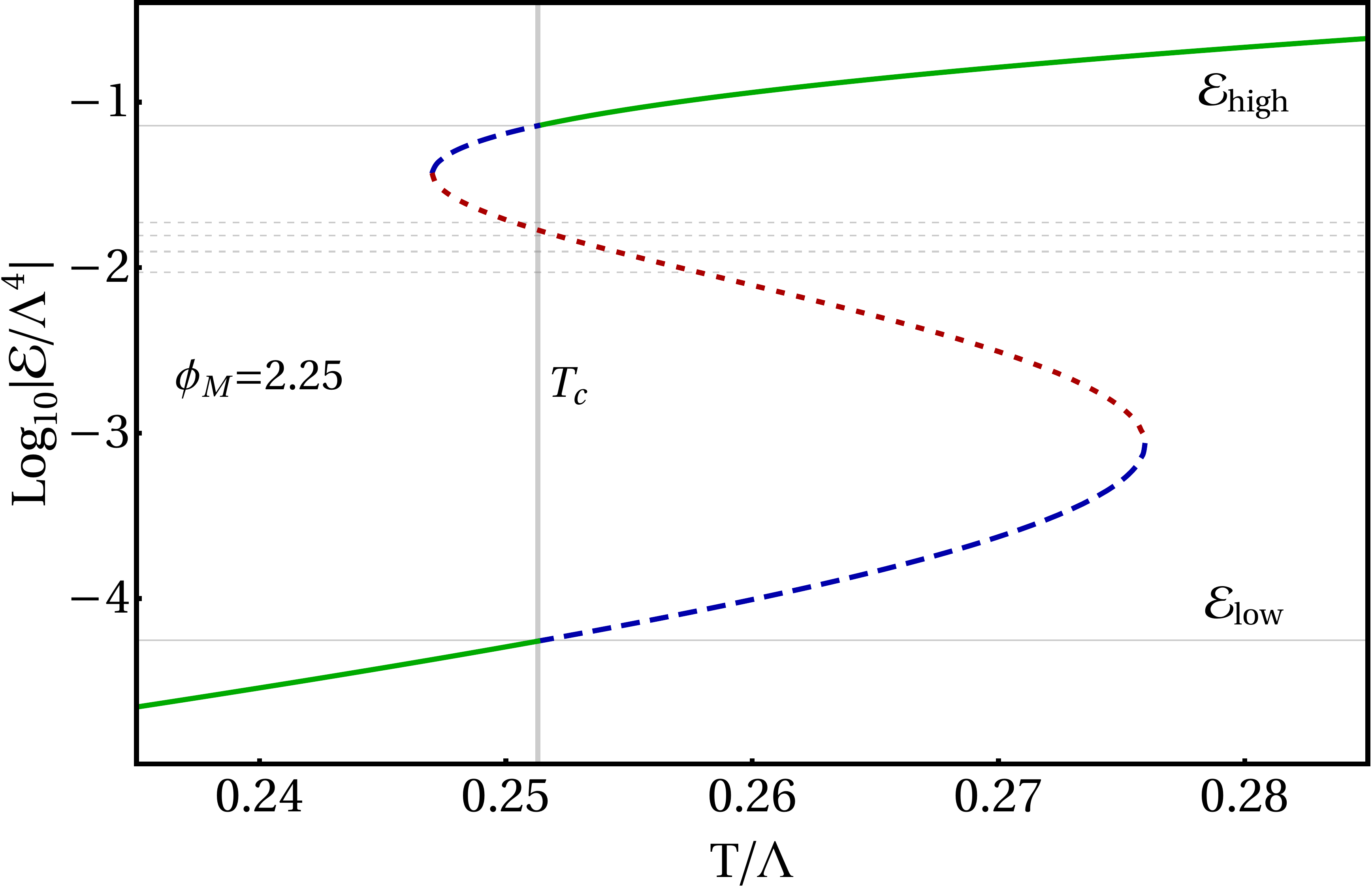}
&
\includegraphics[width=.49\textwidth]{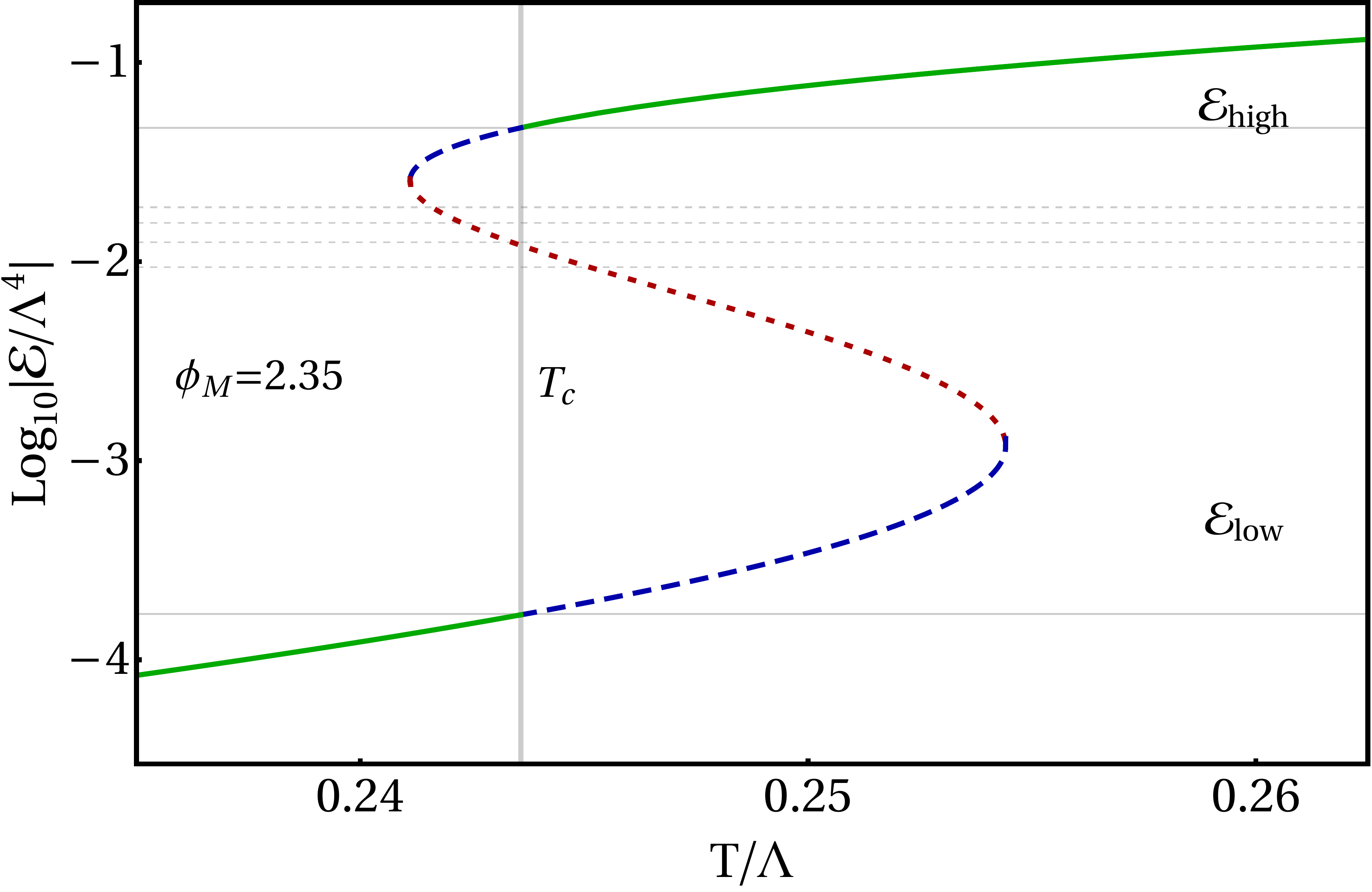}\\[0.4cm]

\includegraphics[width=.49\textwidth]{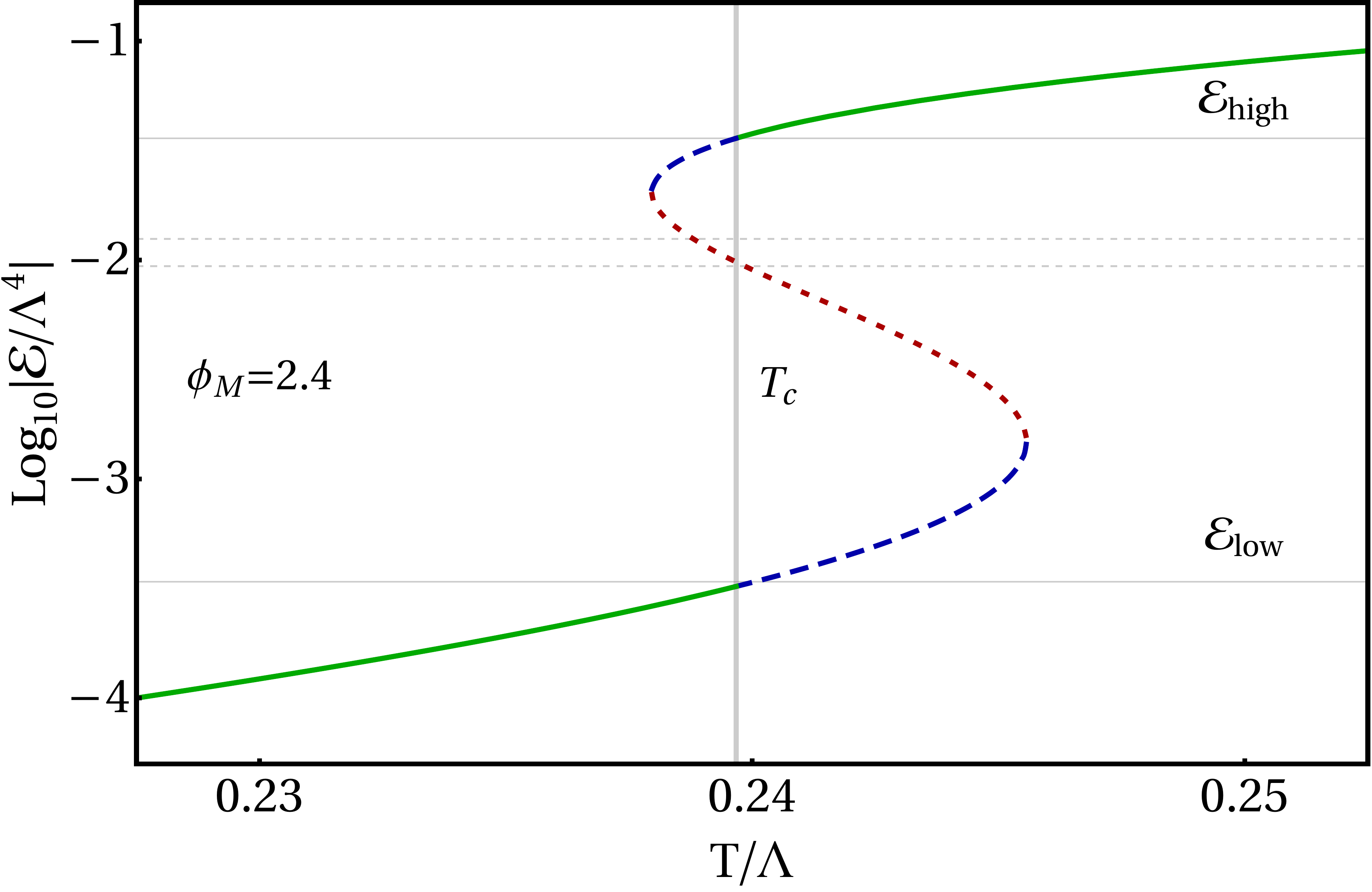}
	&
\includegraphics[width=.49\textwidth]{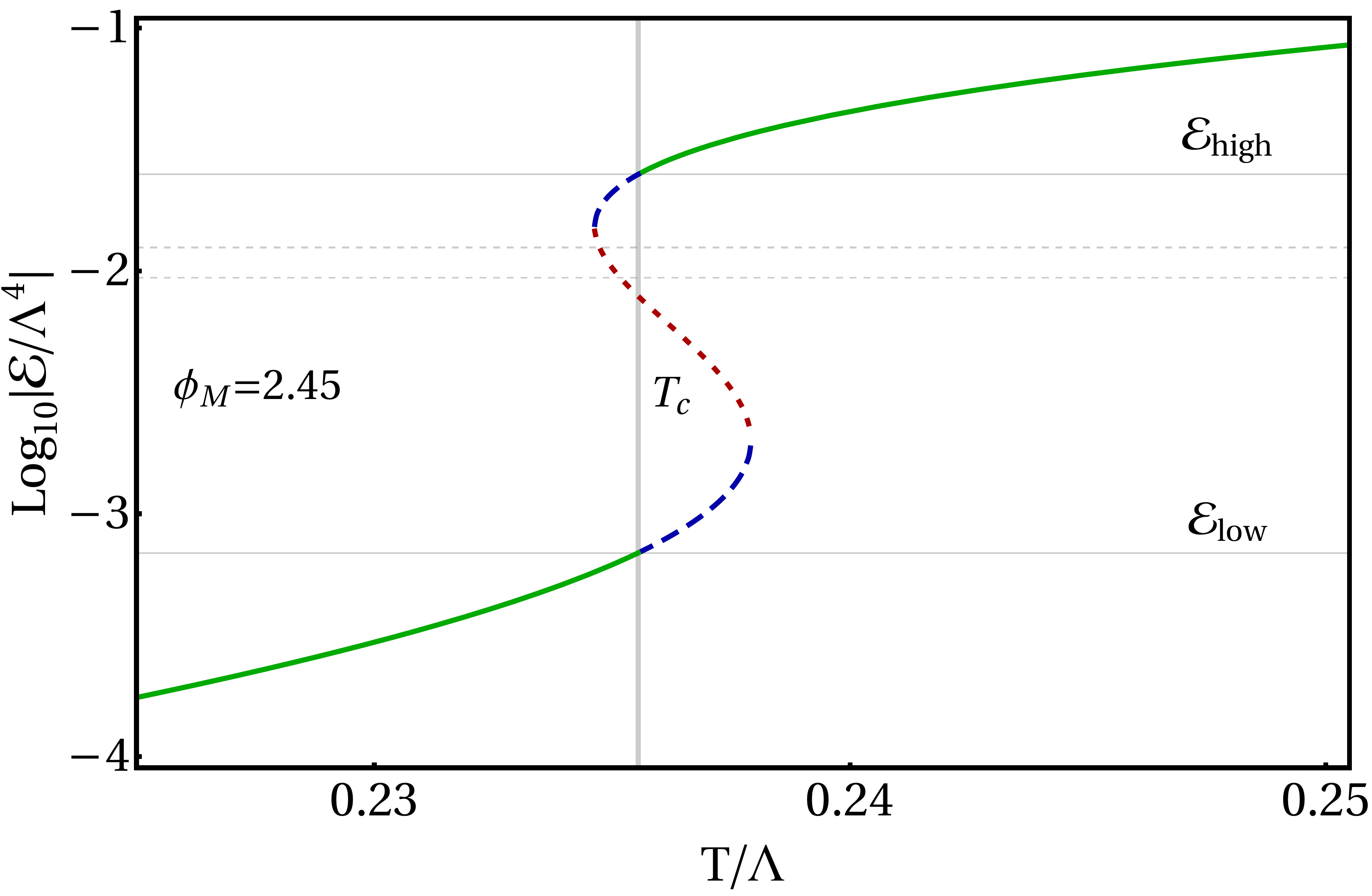}
\end{tabular}
\end{center}
\vspace{-5mm}
	\caption{\label{equationofstates}Energy density versus temperature for the theories with different first order phase transitions with varying criticality due to different parameter $\phi_{\rm M} = \{2.25, 2.35, 2.4, 2.45\}$. The full green line represents the energy densities in the cold and hot stable phases, the dashed blue in the meta stable regions and the dotted red in the unstable region. The vertical full gray line marks the temperature of the phase transition $T_c$. The horizontal full grey lines mark the stable energy densities $\Elow$ and $\Ehigh$ at the transition temperature. The horizontal dashed grey lines are the four initial energy densities $\{\E_1 , \E_2, \E_3, \E_4\}$ or just the two initial respective $\{\E_1 , \E_2 \}$ of the simulations.}
\end{figure*}
Each theory set by a different value of \eqref{eq:phimvalues} has a first order phase transition.
The strongest phase transition (farthest from a critical point) in this setup occurs for $\phi_{\rm M} = 2.25$,
while the softest (nearest to a critical point), has $\phi_{\rm M} = 2.45$ as one approaches the theory with critical point $\phi_{\rm M}^* \approx 2.521$.
\footnote{It is numerically not practicable to get much closer to the critical point as due to the small range of unstable modes the spinodal instability then needs exponentially longer to kick in.}
All the considered states are deconfined similar to the characteristic of the quark-gluon plasma.
Note that the parameter $\phi_{\rm M}$ always appears as a square in the potential and therefore also in the equations of motion.
Hence a small change in $\phi_{\rm M} \leq \phi_{\rm M}^*$ results in quite different theories and correspondingly in different equations of state.\\
In the trace of the stress tensor
\begin{align}
\label{eq:TTrace0}
\left<T^{\mu}_\mu\right>= - \Lambda \left< \mathcal{O} \right> \,,
\end{align}
one recognizes the scale $\Lambda$, which sets the magnitude of the non-normalizable mode of the scalar field and is the source of the conformal invariance breaking.
The rescaled stress tensor given by
\begin{align}
({\cal E}, P_L, P_T, \mathcal V )=
 \tfrac{\kappa_5^2}{2 L^3} (-T^t_t, T^z_z, T^{x_\perp}_{x_\perp}, \mathcal O )\,,
\end{align}
omitting the expectation value signs,
introduces the local energy density $\cal E$, the longitudinal pressure $P_L$, the transverse pressure $P_T$ and the expectation value of the scalar operator $\mathcal V$. Here $z$ is the dynamical and longitudinal direction, while $x_\perp$ are the transverse infinite homogeneous directions.\\
For the theories with values of $\phi_{\rm M}$ given in \ref{eq:phimvalues}, I compute
the relevant thermodynamical quantities. Fig.~\ref{equationofstates} shows
the energy density versus temperature with the additional indications of the
local energy density of the stable high phase $\Ehigh$, the stable low energy
phase $\Elow$, the transition pressure $P_c$ and the transition temperature
$T_c$
(the thermodynamics of the theory with $\phi_{\rm M} =2.3$ has already been calculated~\cite{Attems:2017ezz,Attems:2019yqn}, so it is omitted in Fig.~\ref{equationofstates}).
As seen in Fig.~\ref{equationofstates} the thermodynamics depend crucially on
the parameter $\phi_{\rm M}$ and each choice illustrates a different
first-order phase transition.
Going from $\phi_{\rm M} =2.25$ to $\phi_{\rm M} =2.45$, one notices the
first order phase transition to become smoother and less pronounced.
This results in the shrinking of the unstable region in
Fig.~\ref{equationofstates}, plotted in dashed red in the equation of state,
both in the temperature and the local energy density range.

\begin{figure*}[t]
\begin{center}
\begin{tabular}{cc}
\includegraphics[width=.49\textwidth]{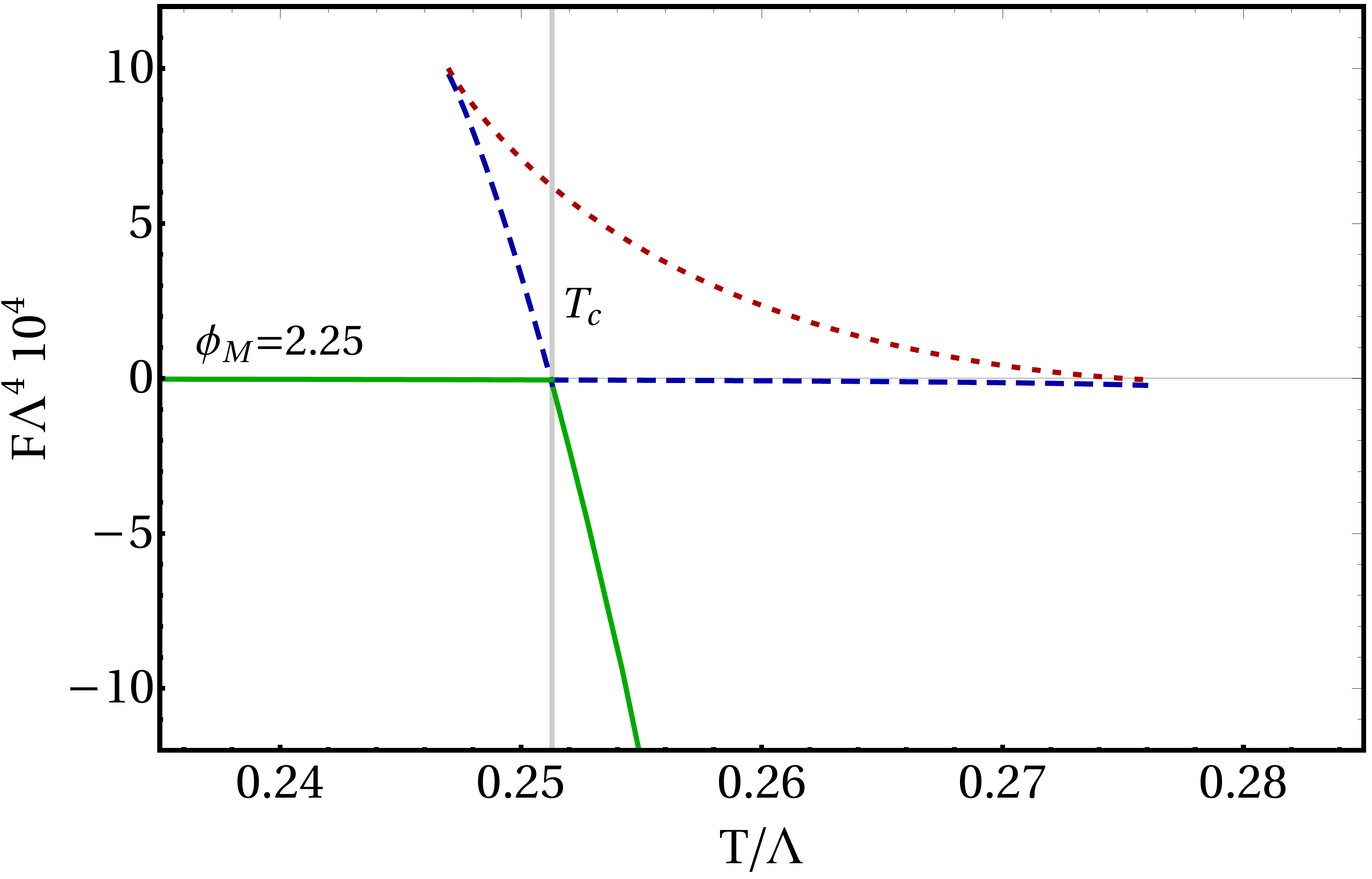}
&
\includegraphics[width=.49\textwidth]{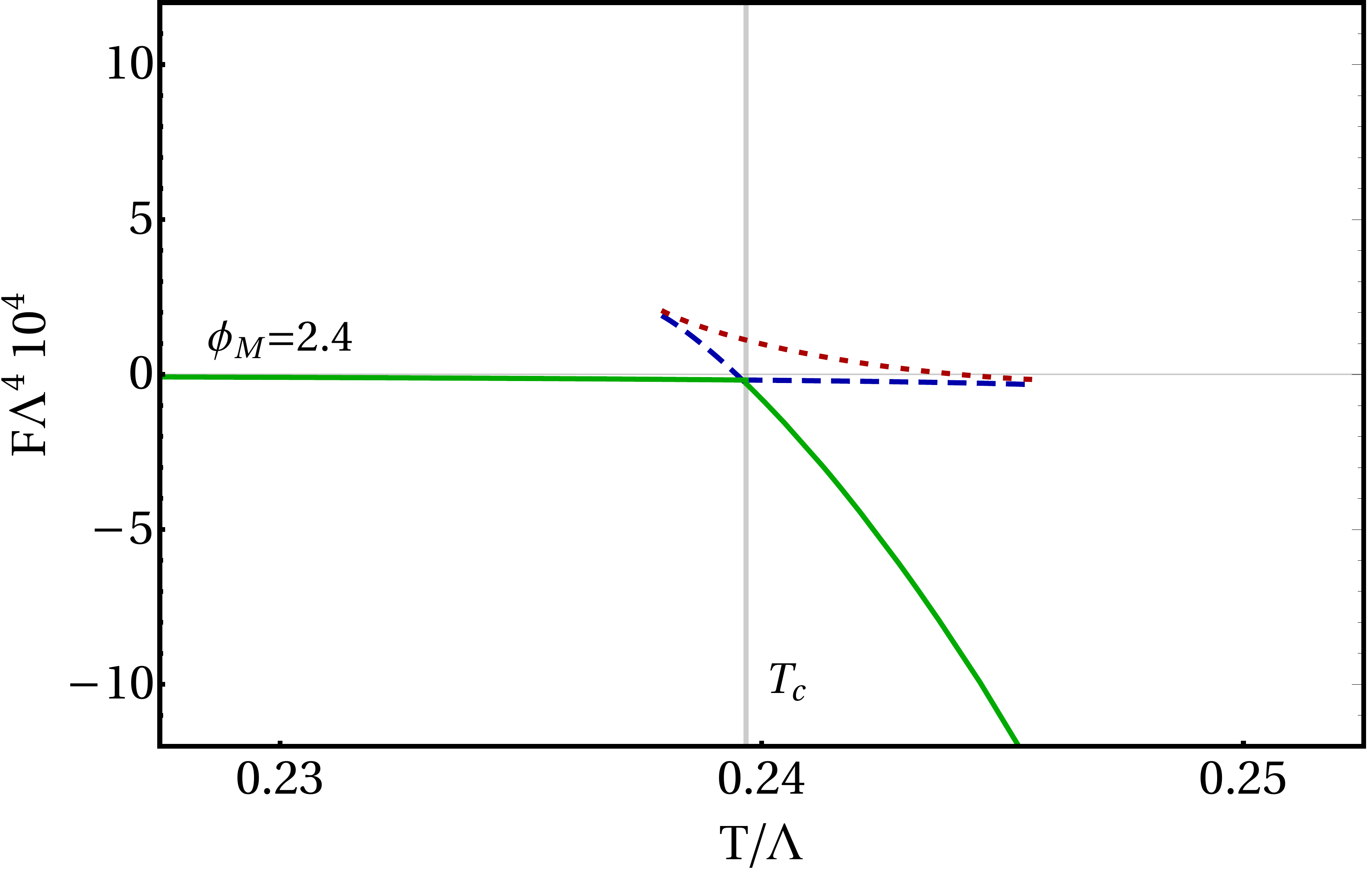}\\[0.4cm]
\includegraphics[width=.49\textwidth]{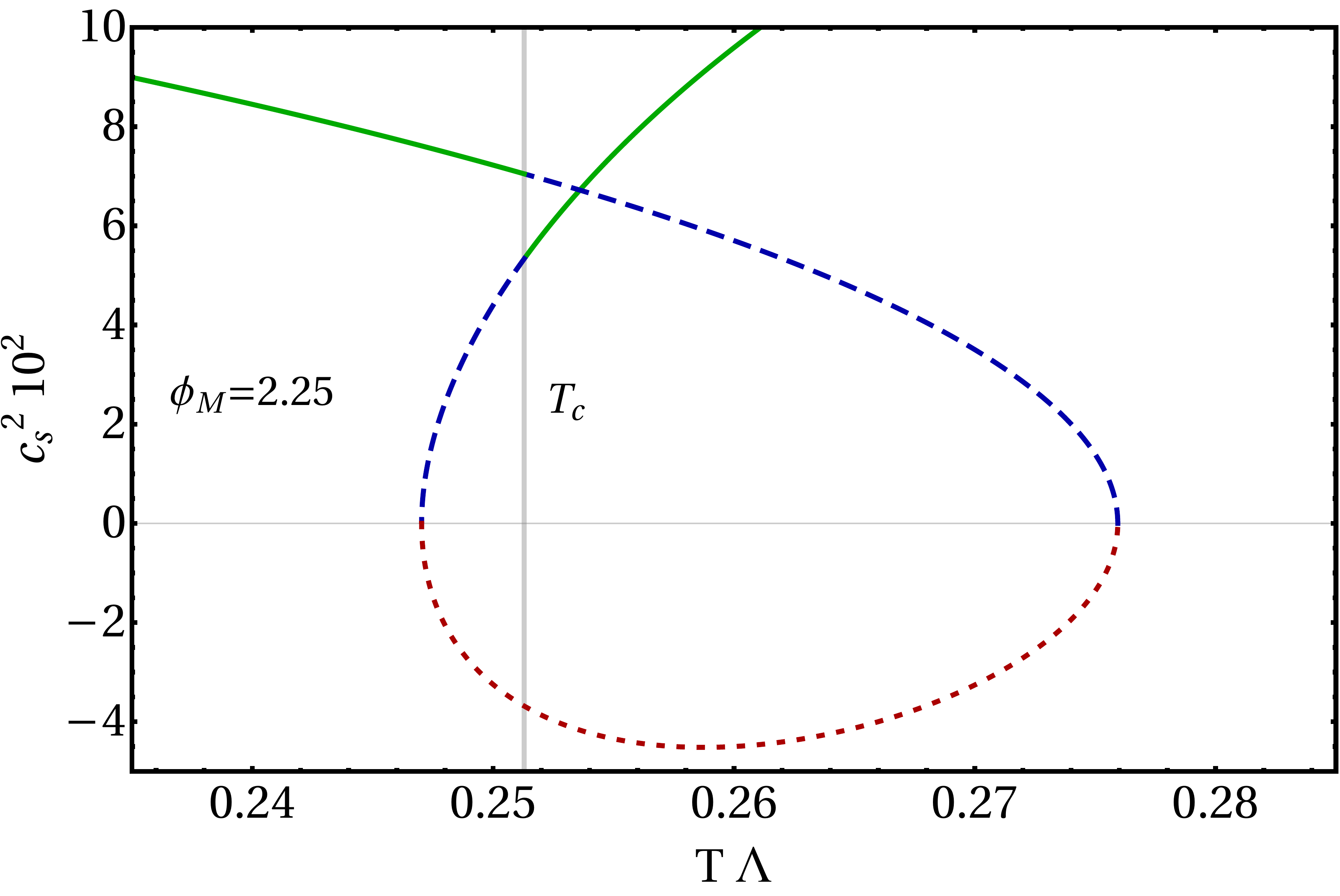}
&
\includegraphics[width=.49\textwidth]{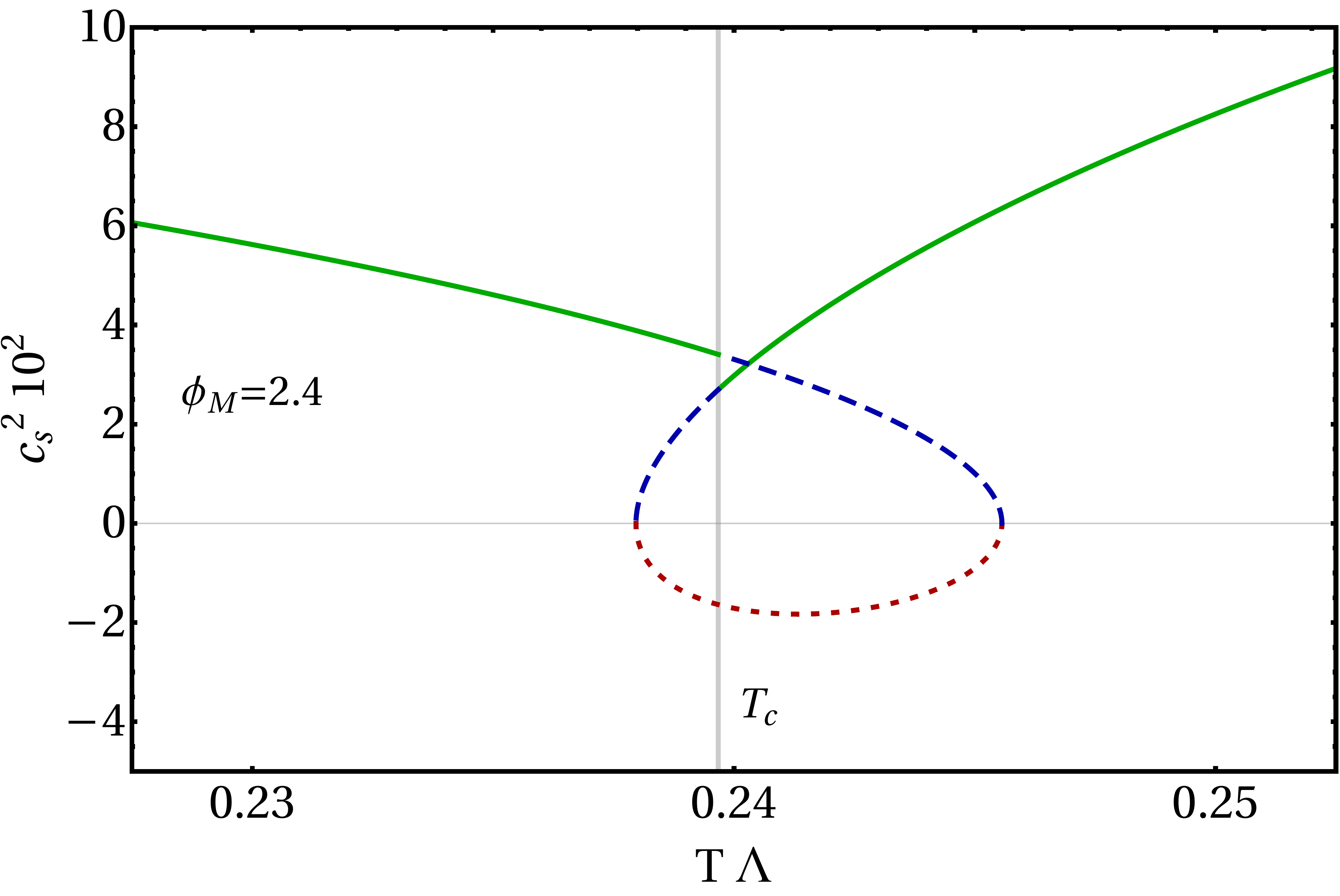}
\end{tabular}
\end{center}
\vspace{-5mm}
	\caption{\label{freeenergyandcssquared}The free energy $F$ and the speed of sound squared $c_s^2$ over temperature for the theories with different first order phase transitions far from the critical point and closer to the critical point with varying criticality due to different $\phi_{\rm M} = \{2.25, 2.4\}$. Each phase is denoted by the same color scheme as in Fig.~\ref{equationofstates}. }
\end{figure*}
In Table~\ref{tab:crit} are listed the respective $\Ehigh$, $\Elow$, $P_c$ and
$T_c$ for each theory with varying first-order phase transition.
In the section \ref{sec:crit}, I will use the pressure of the transition $P_c$
to determine the difference between the inhomogeneous states.
At the transition temperature $T_c$ homogeneous stable states have
the same equilibrium pressure $P_c$.
There are several ways to compute the transition temperature $T_c$.
One way is via the crossing of the free energy $F$ as illustrated in the top
plots of Fig.~\ref{freeenergyandcssquared}.
For the energy densities in the unstable region between $\Elow$ and $\Ehigh$ - the spinodal region,
the speed of sound squared is negative as plotted in the lower plots of
Fig.~\ref{freeenergyandcssquared} and the states are affected by a long-wave
length instability.
\begin{table}[h!]
\begin{center}
\caption{Local energy densities for the stable cold and hot phases of each theory\\and the corresponding transition pressures and temperatures.}
\label{tab:crit}
\begin{tabular}{l|l|l|l|l|l}
      \toprule
	    $\boldsymbol{\phi_{\textrm M}}$ & $2.25$ & $2.3$ & $2.35$ & $2.4$ &$2.45$\\[0ex]
      \midrule
    $\boldsymbol{\Ehigh /\Lambda^4}$ & $7.2 \times 10^{-2}$ &$5.9 \times 10^{-2}$ & $4.7 \times 10^{-2}$ & $3.5 \times 10^{-2}$ & $2.5 \times 10^{-2}$\\
 $\boldsymbol{\Elow /\Lambda^4}$ & $5.6 \times 10^{-5}$ & $9.4 \times 10^{-5}$ & $1.7 \times 10^{-4}$ & $3.5 \times 10^{-4}$ &$6.9 \times 10^{-4}$\\
    $\boldsymbol{P_c /\Lambda^4}$ & $0.3 \times 10^{-7}$& $7.5 \times 10^{-6}$ & $0.7 \times 10^{-6}$ & $1.1 \times 10^{-6}$ &$2.9 \times 10^{-5}$\\
   $\boldsymbol{T_c /\Lambda}$ & $0.251$ & $0.247$ & $0.244$ & $0.240$ & $0.236$\\
   \bottomrule
    \end{tabular}
  \end{center}
\end{table}
Note the stark difference between the two stable local energy
densities from more than three to less than two orders of magnitude.
Due to the more than three orders of magnitude difference between the stable
cold and hot phases and due to the large spikes in gradients during the
non-linear regime of the spinodal instability, the numerical real-time treatment
of this system is extremely challenging.
Between $\phi_{\rm M} =2.25$ and $\phi_{\rm M} =2.45$ the value for the hot
stable phase $\Ehigh$ more than halves and the value for the cold stable phase
$\Elow$ is approximately multiplied by twelve.
The change in $\Ehigh$  is significant as the system with $\phi_{\rm M} =2.25$  compared to $\phi_{\rm M} =2.45$  needs more than twice as much
total integrated energy density for a fully phase separated setup.
Similarly, the change of  magnitude in $\Elow$ is significant as it sets
the behaviour of the phase domain wall. Moreover, it eases or hardens the
numerical treatment.
Obviously, there is a trade off between the needed numerical resolution for the stronger instability and the much longer evolution of the softer instability.

The setup uses homogeneous planar black brane solutions periodic in the $z-$direction with a finite box extent $L_z$ and different initial local energy densities:
$\E (t = 0) =\{\E_1 , \E_2 , \E_3, \E_4\}$.
For the softer theories with  $\phi_M = \{2.4, 2.45 \}$ and a smaller $\Ehigh$,
it is sufficient to simulate setups with $\E (t = 0) =\{\E_1 , \E_2 \}$.
Moreover, for the theory with the parameter  $\phi_M = 2.45$ the bigger
energy densities $\E_3$ and $\E_4$ are no longer in the unstable phase:
\begin{align}
	\{\E_1 , \E_2 , \E_3, \E_4 \} = \{ 0.9, 1.2, 1.6, 1.9 \}\times 10^{-2} {\Lambda^4} \,.
\end{align}
	In order to trigger the spinodal instability, one only has to slightly perturb these homogeneous solutions at the start of an evolution or wait for the numerical white noise to kick-in.
Here I use a small sinusoidal perturbation with the amplitude of $\Delta = 10^{-4}$ which, thanks to non-linear coupling, populates all unstable modes.
This is much faster than numerical noise which needs to build up the unstable
modes from amplitudes below $10^{-12}$.
All the boundary data of the simulations is published for further analysis or comparisons as open data on a Zenodo repository~\cite{attems_maximilian_2020_3445360}.
As all the chosen initial local energy densities are in the dynamical unstable
region between $\Elow$ and $\Ehigh$ of each theory, the plasma will be subject
to the spinodal instability, whose dynamical properties one is interested in.

\section{Criterium for inhomogeneous states}\label{sec:crit}
The spinodal instability cools down regions in the unstable energy density, so
that they reach the cold stable phase and hence pushes out the energy density
to other regions, which in turn accumulates and may reach the hot stable phase.
This results in various inhomogeneous states, which are to be further classified.
In what follows I will define the two distinct maxima and minima that
form due to the spinodal instability: a peak and a plateau;
a gorge and a valley.
\subsection{Definition: peaks versus plateaux}
While the maxima domains, peaks versus plateaux, can be respectively defined by
their total energy density content and the typical order of magnitude of
difference between them, there is a more distinct and stringent criterium:
the transverse pressure $P_T$.
When the maximum of the transverse pressure at the interface is of the value
of the transition pressure $P_c$, which is the equilibrium pressure
at the phase transition, the inhomogeneous state is a plateaux.
For the criterium the numerical tolerance is set to $20\%$.
\footnote{This is the usual tolerance criterium for the hydrodynamization
process (i.e. the time when hydrodynamics applies).
It can be stretched to $30\%$ or reduced to $10\%$ without much interpretational
change.}
Previously, to classify the simulation one had  to integrate the local
energy density of the state over spacetime and checking both the maximum and
the two minima on the side of the state.
With this criterium, one only needs to extract a single value.
The criterium makes use of the homogeneous spatial direction, which in our case
of planar black branes are the two perpendicular directions $x_\perp$,
to distinguish between the two types of maxima and minima.
The maximum of the transverse pressure of a peak is well below the
critical pressure $P_c$.
Therefore a peak is a local maximum.\\

Note that in contrast to the transverse pressure $P_T$,
the longitudinal pressure $P_L$ is not a useful thermodynamical quantity
for differentiating between the inhomogeneous states
as it correlates with the fluid velocity of the formed peak or plateau.
The longitudinal $z$ direction is the dynamical direction of the simulations.
Therefore $P_L$ is very useful to check how well settled the ongoing simulation is.
Once there is no ongoing dynamics, the longitudinal pressure equals up to
numerical precision the critical pressure at any position $z$ on the boundary
of the spacetime $P_L (z, t > t_{\rm final}) = P_c$.
At late times the fluid velocity vanishes, as one reaches the static
configuration.
This means the longitudinal pressure is constant up to numerical precision once the system is settled.  \\

For the two minima domains, the gorge and the valley, the same
definition as for the maxima applies.
The valley, in analogy to an extended U-shaped formation,
has its transverse pressure at the critical pressure $P_c$.
While the gorge, in analogy to a narrow V-shaped minima formation, has
a transverse pressure lower than the critical pressure.
The gorge corresponds to a local minima in the local energy density above
$\Elow$.
In all the simulations the gorge state is rarely realized.
It happens intermittently before the merger of peaks or domains, but rarely
forms in the reshaping stage as the spinodal instability tends to cool down
to the lower stable phase forming valley extents.\\

The presence of a plateaux enforces a phase separated interface,
where the transverse pressure inside the plateaux attains the critical
pressure $P_c$.
On the contrary, the maxima of the transverse pressure  of a  peak does not
reach  $P_c$ and its local energy density might not reach the hot
stable phase.
The distinction is significant and easy to compute by extracting
the corresponding maximum of the transverse pressure.
The distinction between peaks and plateaux is of course fluid, since a peak
with sufficient local energy density input can be turned into a plateaux.
Nevertheless, this new criterium greatly facilitates their distinction and
is of interest for the study of the properties of the phase separation
especially important as only plateaux have the full phase separation.\\

\subsection{Final stage}
\begin{figure*}[t]
\begin{center}
\begin{tabular}{cc}
\includegraphics[width=.51\textwidth]{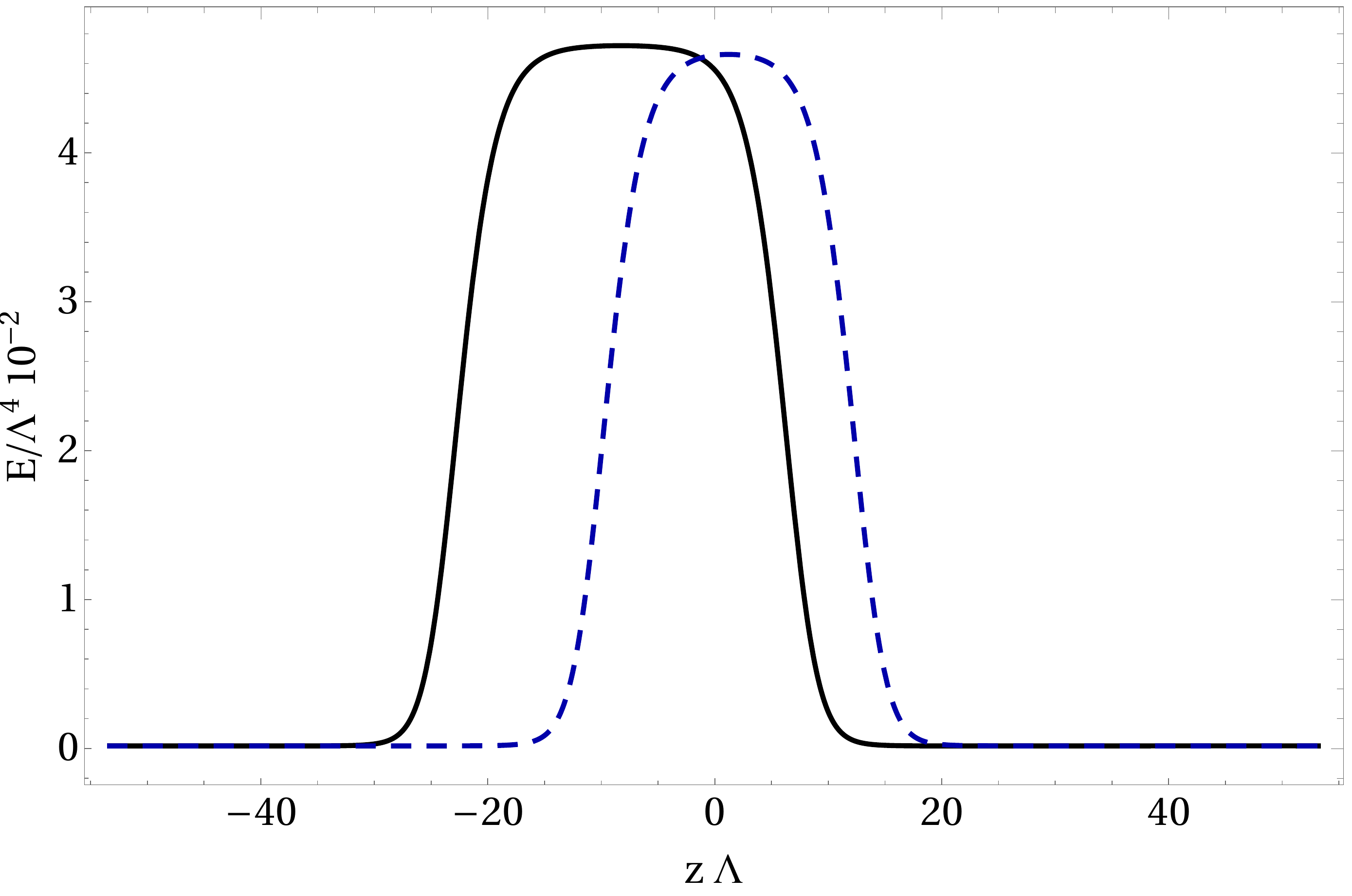}
\quad \hspace{-2mm}
\includegraphics[width=.51\textwidth]{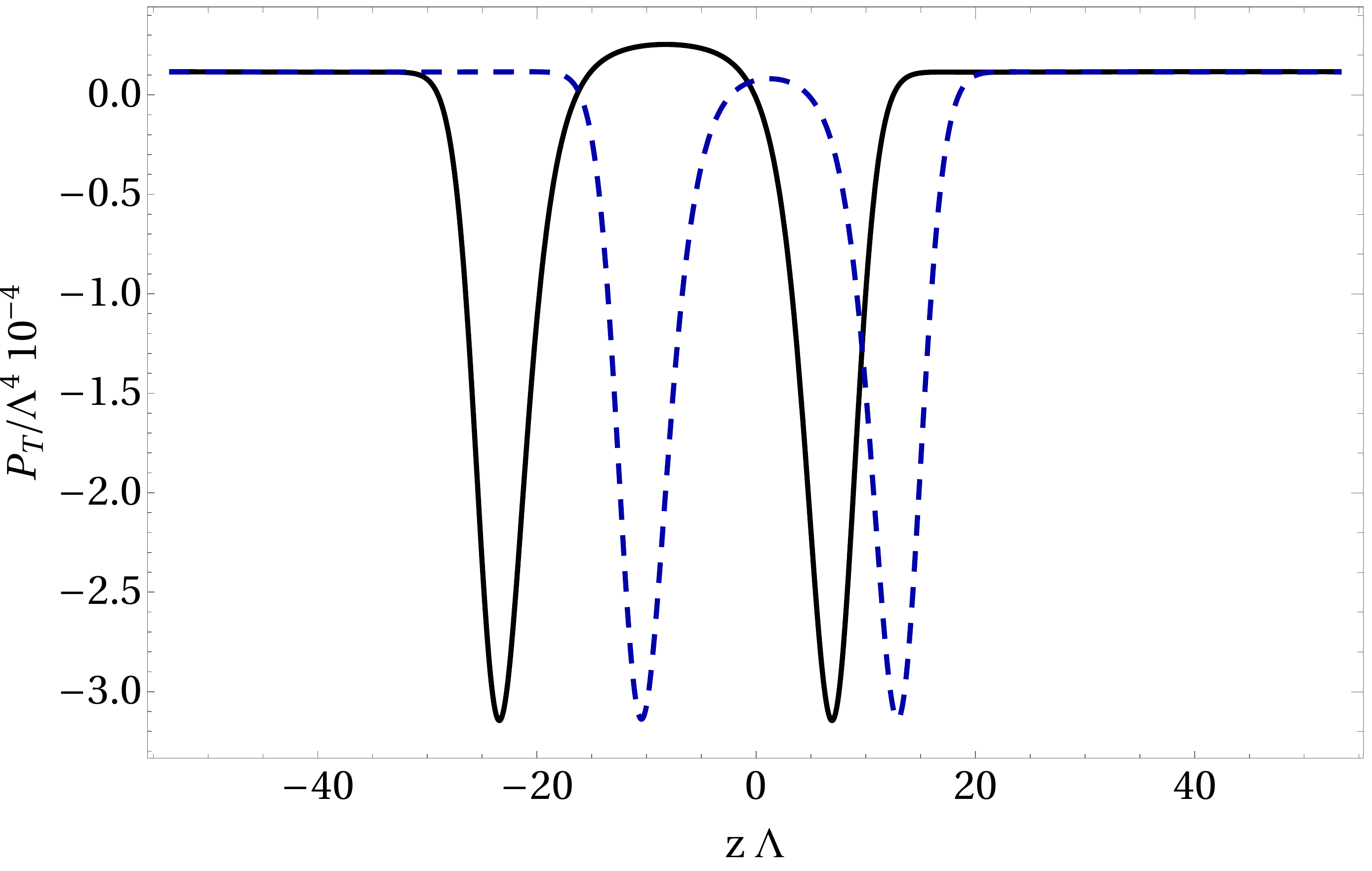}
\end{tabular}
\end{center}
\vspace{-5mm}
\caption{\label{fig:domainvspeak} Longitudinal superposed profiles with $\phi_{\rm M} = 2.35$ and $L_z \Lambda = 107$ for a single plateau in continuous black with initial energy density $\E_2$  and for a single peak in dashed blue with initial energy density $\E_1$ of
	(left)
	the local energy density
	(right)
	the transverse pressure
	}
\end{figure*}
If the simulated box contains enough total energy density,
the final stage of the spinodal instability will be
an inhomogeneous phase separated solution
(neglecting for now subtleties of quite small finite boxes,
where finite size effects act as regulator of the instability~\cite{Bea:2020ees}) that forms a plateaux.
In the special case of not enough total energy density in the system the formed
final stage is a single peak.\\

The transverse pressure of the final stage of a simulation with the initial
homogeneous local energy density $\E(t=0) = \E_2$ is plotted in the black
continuous curve of Fig.~\ref{fig:domainvspeak} (right).
The maximum value of the transverse pressure $P_T(z \approx -4.1) \approx 11.0 \times 10^{-5}$ is largely within the tolerance criterium of the critical pressure ($\approx 95\% P_c$),
therefore it is a plateau.
Whereas the maximum of the transverse pressure of the final stage  with the
same theory but less initial local energy density $\E(t=0) = \E_1$ has a
value of around $~70\% P_c$ below the tolerance criterium, therefore its final
inhomogeneous state is a peak.
The final longitudinal profile of the transverse pressure is plotted in the
blue dashed curve in Fig.~\ref{fig:domainvspeak} (right).
The maximum in $P_T$ always corresponds to the location
at the longitudinal direction $z$
of the maximum of the local energy density $\E$.
As seen in Fig.~\ref{fig:domainvspeak} (left) the plateau has clearly a large
extent near the hot stable phase in the local energy density, while the peak
only touches it briefly around its maximum.
In both cases (black continuous and blue dashed) the extended minima
Fig.~\ref{fig:domainvspeak} (right) has the value of the critical pressure and
hence it is a valley.
Consequently the simulation in full black is a neat example of
a fully phase separated solution.\\

\subsection{Reshaping stage}\label{subsec:reshaping}
\begin{figure*}[t]
\begin{center}
\begin{tabular}{cc}
\includegraphics[width=.51\textwidth]{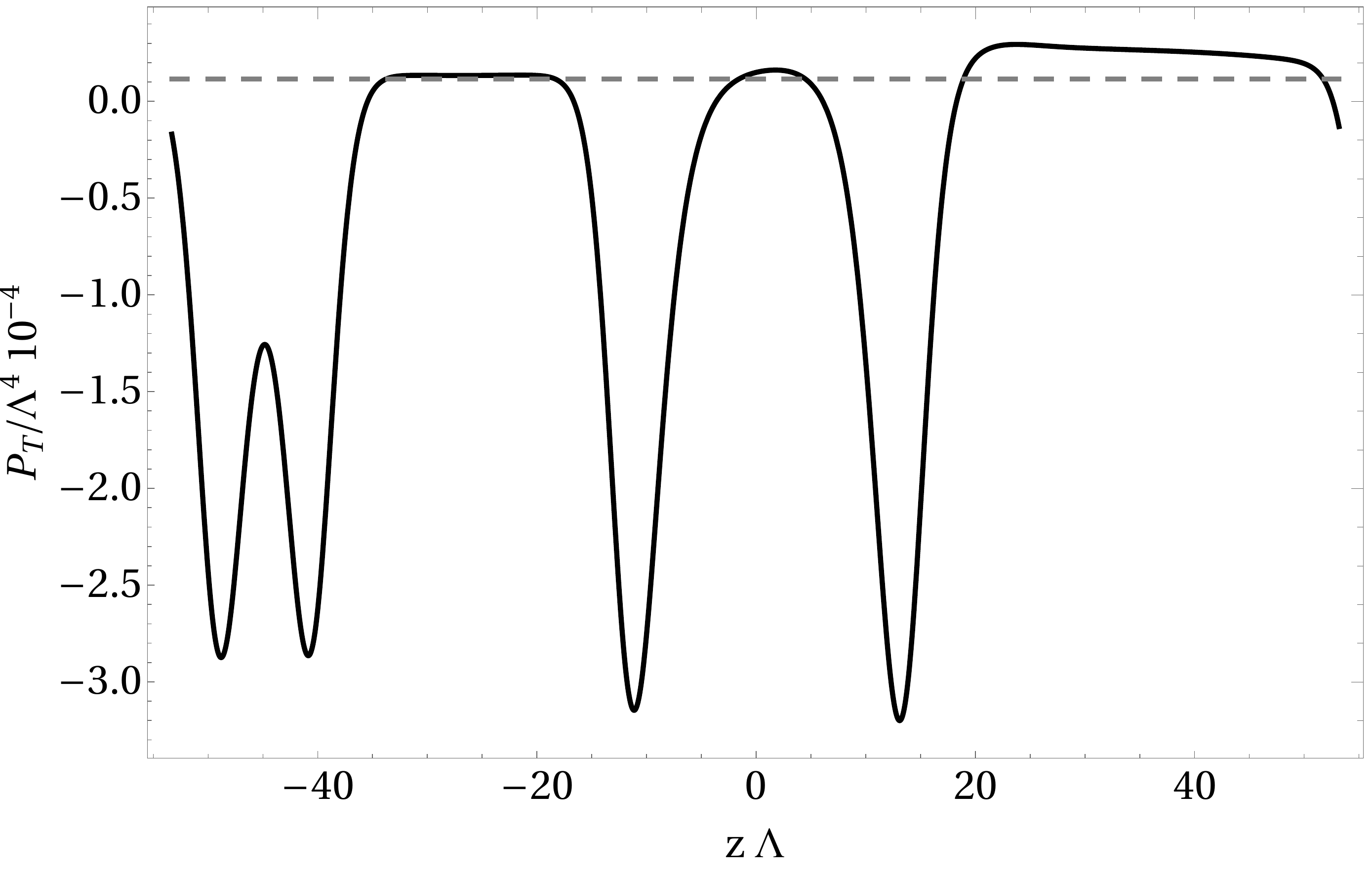}
\quad \hspace{-2mm}
\includegraphics[width=.51\textwidth]{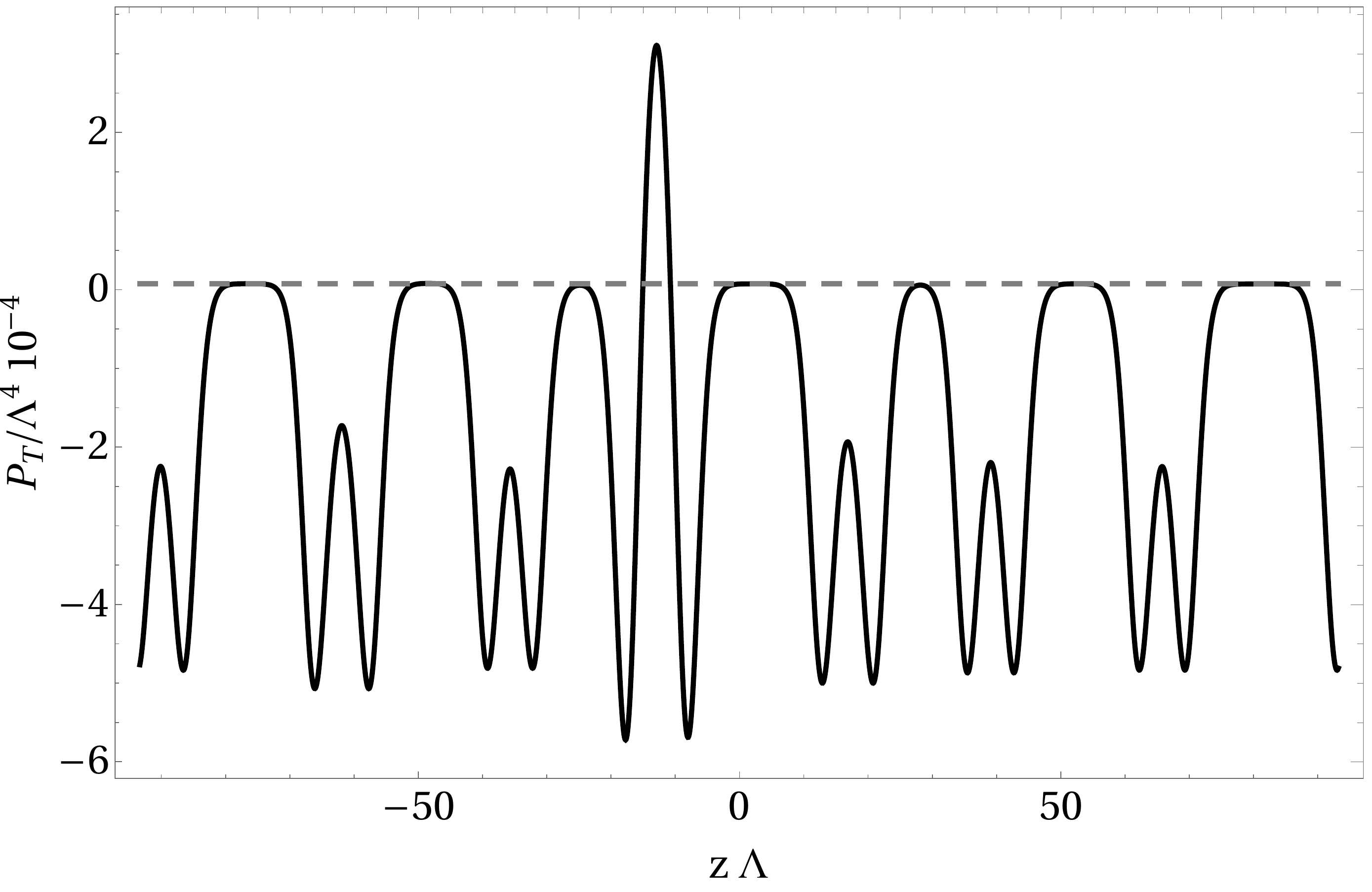}
\end{tabular}
\end{center}
\vspace{-5mm}
	\caption{\label{ptcriteriumreshaping} Transverse pressure $P_L$ longitudinal profile during reshaping as the solid black line and the critical pressure $P_c$ as the dashed gray line  (left) with $\phi_M = 2.35$ at $t \Lambda = 1200$ forming a peak, two valleys and a plateau of the simulation visualized on bottom left of Fig.~\ref{plateauformationquad}
	(right) with $\phi_M = 2.3$ at $t \Lambda = 1180$ forming several peaks, a single plateau,  and valleys in between of the simulation visualized on the left Fig.\ref{mergers} }
\end{figure*}
The plateaux are either formed directly early on in the reshaping stage
or later on during the merger stage from several joined peaks.
Again the longitudinal extent of the simulations is assumed to be wide
enough to at least fit a phase separated system with two interfaces.
Here are given examples of directly formed peaks and plateaux.\\

Of course in the early reshaping stage the distinction is by definition a lot
messier compared to the settled final stage due to the ongoing dynamics.
Nevertheless it is possible to distinguish between peaks and plateaux
based on their transverse pressure with the relaxed criterium by allowing
the maximum of a plateau or a valley to overshoot the equilibrium transition
pressure.
For example in Fig.~\ref{ptcriteriumreshaping}(left) one recognizes a single
peak at the position $z \Lambda \approx -45$ and a single plateau at $z
\Lambda \approx 0$ forming.
It also indicates that at this particular moment the valley from
$z \Lambda \approx 20$ to $z \Lambda \approx 52$ is
also still out-of-equilibrium.
The valley between $-35 \lesssim z \lesssim -15$ is much more
equilibrated.
In the simulation with a wide extent shown in
Fig.~\ref{ptcriteriumreshaping}(right) one recognizes the formation of seven
valleys, a single plateau at the position $z \Lambda \approx -12$ and six
peaks, where the transverse pressure maximum is far lower than the transition
pressure $P_c$.
During its reshaping the maximum value of a plateau will severely overshoot the
transition pressure $P_c$ while the maxima of the peaks do not reach $P_c$
during their reshaping.

After reshaping in the merger stage, one observes the
peaks to settle quite quickly and have only small changes over time:
The maximum energy density of a peak decreases by less than $<1\%$ over
very long times\cite{Attems:2019yqn}.
Relatedly the transverse pressure changes by the same amount
over long period times.
In the merger stage the formed peaks typically merge together to plateaux.
Whereas plateaux take a long time to settle, the more extended the longer and
asymptotically tend to $P_c$.
Therefore the relaxed inhomogeneous criterium works equally well during
the reshaping and merger stages:
By checking the transverse pressure of the inhomogeneous states to be
far above or at least attaining a transverse pressure bigger than the
tolerance criterium ($\ge 20\% P_c$) one is able to handily distinguish
plateaux and peaks.

\subsection{Merger stage}
\begin{figure*}[t]
\begin{center}
\begin{tabular}{cc}
\includegraphics[width=.51\textwidth]{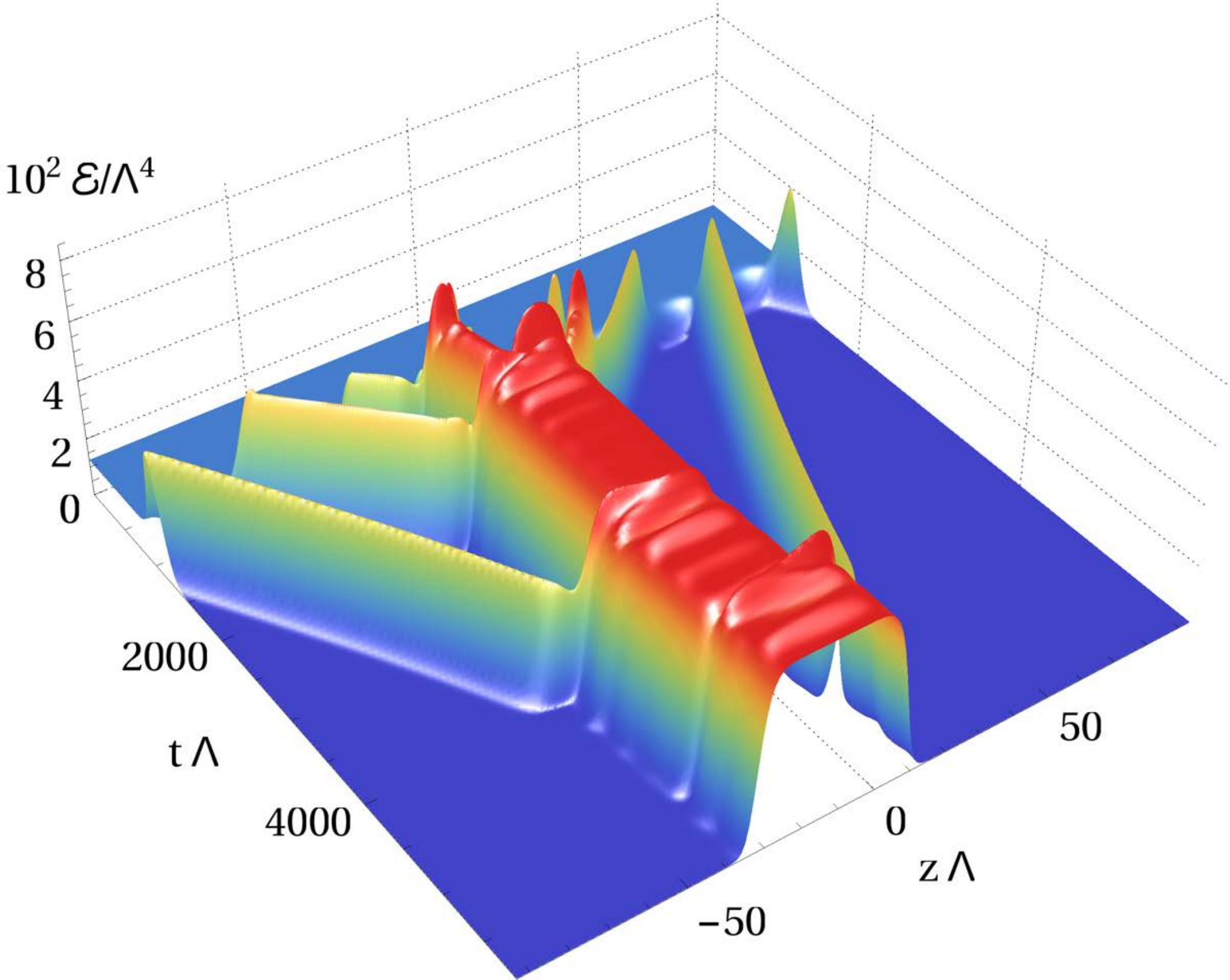}
\quad \hspace{-2mm}
\includegraphics[width=.51\textwidth]{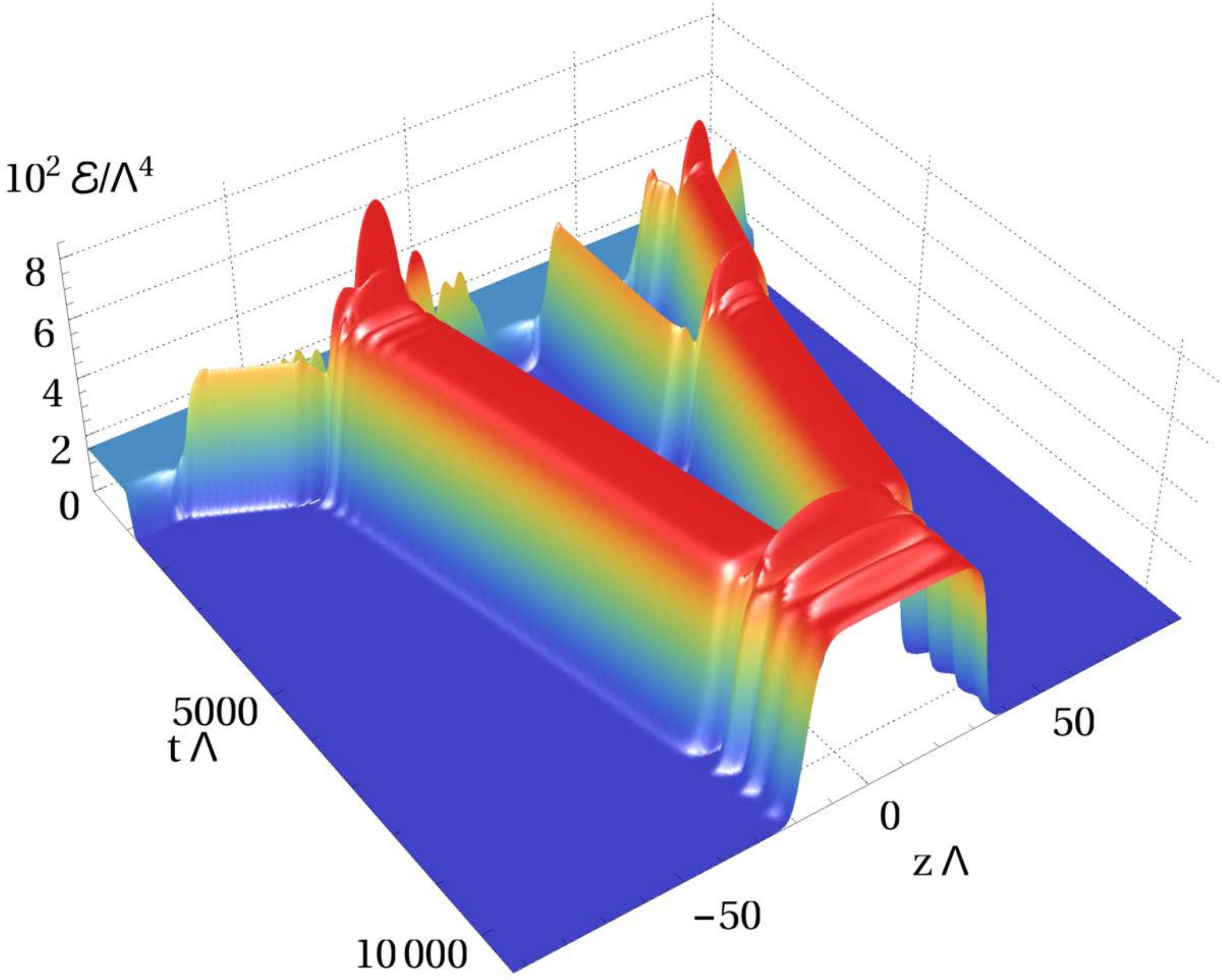}
\end{tabular}
\end{center}
\vspace{-5mm}
	\caption{\label{mergers} Time evolution of two different merger with $\phi_{\rm M} = 2.3$ and $L_z\Lambda \simeq 187$ longitudinal extent: (left) Spacetime evolution $\Ezero=\E_1$ and with initial $n=1$ perturbation up to $t\Lambda = 5885$; (right) Spacetime evolution $\Ezero=\E_2$ and numerical noise as initial perturbation up to $t\Lambda = 10700$.}
\end{figure*}

An even more challenging stage for the distinction criterium is the
merger stage.
In the previous subsections the static final and the early reshaping stage
in which each inhomogeneous state form have been discussed.
Due to the static nature of the final stage and the usual crisp formation
of the spinodal instability those stages are easier to classify.
In the so-denominated merger stage  the complicated dynamics with the
peaks and domains merging happens.
Here one discerns that mergers happen in a same qualitative way for
the different theories:
Peak $\leftrightarrow$ peak and peak $\leftrightarrow$ plateau mergers
lead to the preferred final solution of a single phase separated plateau.
It is worth point out that forming states with more than one plateau requires
significant computing resources, due to the large longitudinal extent and the
increased evolution time of this large merger and subsequent equilibration.
Here I analyse for the first time two evolution, where the inhomogeneous
formed states look like out-of-equilibrium or equilibrium plateaux mergers
yet are in fact peak $\leftrightarrow$ plateau mergers.
The two evolution of the mergers in the \fig{mergers} are a difficult
example for the classification of the inhomogeneous states by the criterium.
\\

\begin{figure*}[t]
\begin{center}
\begin{tabular}{cc}
\includegraphics[width=.51\textwidth]{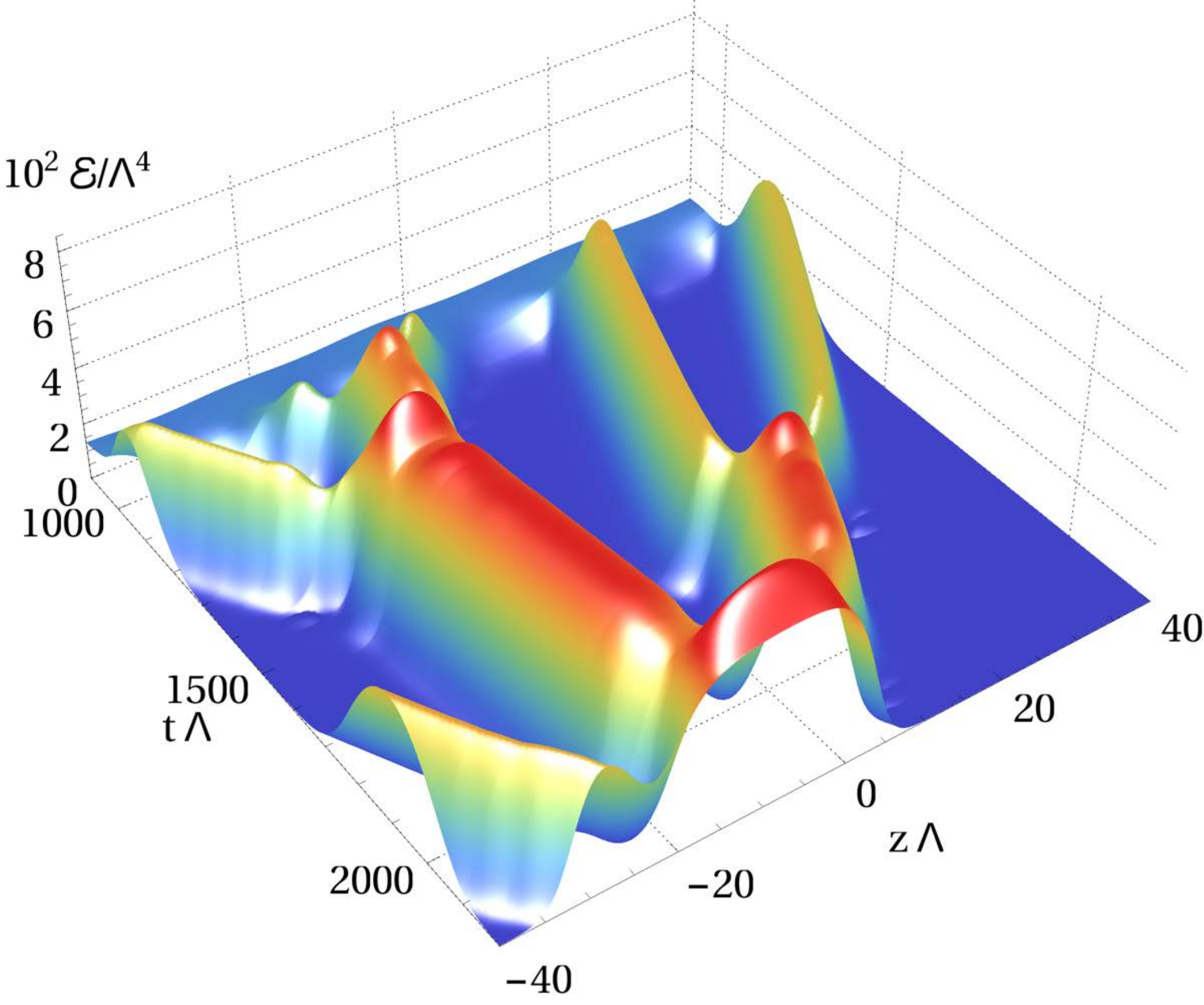}
\quad \hspace{-2mm}
\includegraphics[width=.51\textwidth]{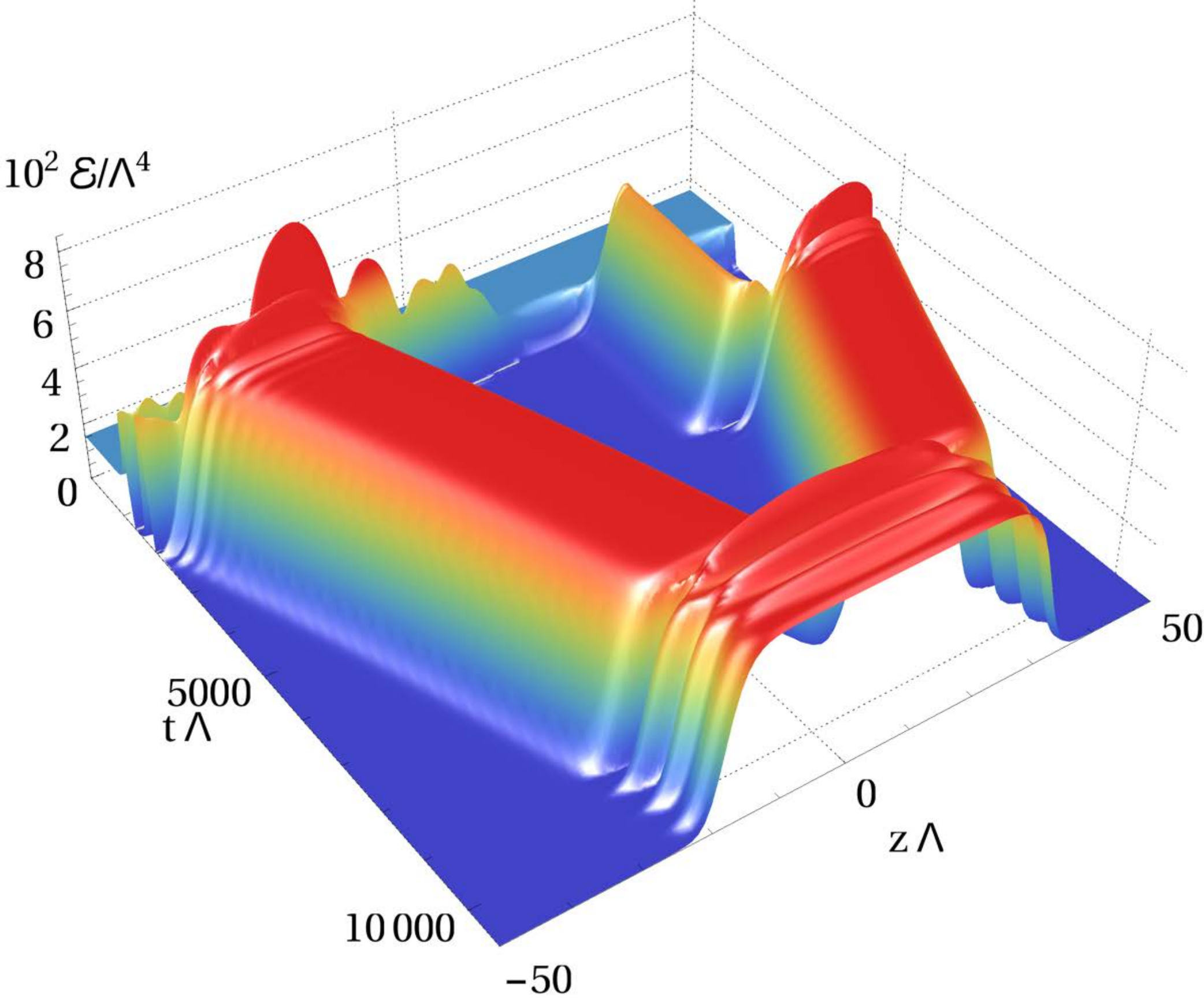}
\end{tabular}
\end{center}
\vspace{-5mm}
	\caption{\label{mergerzooms} Zoom into the respective spacetime evolution of \fig{mergers} (left) out-of-equilibrium peak $\leftrightarrow$ peak merger (right) Merger of settled peak $\leftrightarrow$ domain }
\end{figure*}

In the merger of the \fig{mergers}(left) at around $t \Lambda \approx
2000$, one could be tricked into seeing two plateaux merging,
while the maximum of the left peak only
overshoots early after the left group merges the peaks around $t \Lambda
\approx 1400$ the critical pressure, but distinctively settles below.
The respective maxima each are a product of mergers and they look deceptively
massive enough for being a plateau each.
The right merging group of the \fig{mergers}(left) never comes to
settling between $1700 < t \Lambda < 1900$, but as product of only two
peaks is very likely a peak too.
Hence one can classify in the \fig{mergerzooms}(left) the mergers as being
peak  $\leftrightarrow$  peak.\\

In the merger of the \fig{mergers}(right) at around $t \Lambda \approx
8750$, one could be equally tricked into seeing two $\Ehigh$ plateaux merging.
Here the case is a it easier to disentangle as both maxima pick up almost
constant velocity towards each other, but settle long before the merger.
The left maximum is a plateau, while the right maximum is a peak as its
transverse pressure settles at $\approx 0.75 P_c$.
Consequently the evolution of the \fig{mergerzooms}(right) shows the
process of a peak $\leftrightarrow$ plateau merger.
Previously demonstrated merger in Fig.~13 of \cite{Attems:2019yqn} is
happening in out-of-equilibrium
where each plateaux is still oscillating around $\Ehigh$ before merging.
Therefore it is a third example of a difficult classification during the merger
stage.
The left maximum seems to settle quite below the transition pressure, but
the right maximum around it.
Accordingly it is also a peak $\leftrightarrow$ plateau merger.\\

Hence the new proposed criterium is useful to classify the inhomogeneous
states in all the different stages of the spinodal instability.
Only during subsequent fast violent out-of-equilibrium mergers the state
may not be directly attributable.
As conjectured each evolution reaches the same preferred final state, where the
extent of the ultimate plateau depends on the total energy density of the
calculation.

\section{Characteristics of the interface}
With the clear distinction between different states formed by the spinodal
instability, we will now focus on the characteristics of the phase separation
with varying criticality.
\subsection{Shape}
\begin{figure*}[t]
\begin{center}
\begin{tabular}{cc}
\includegraphics[width=.51\textwidth]{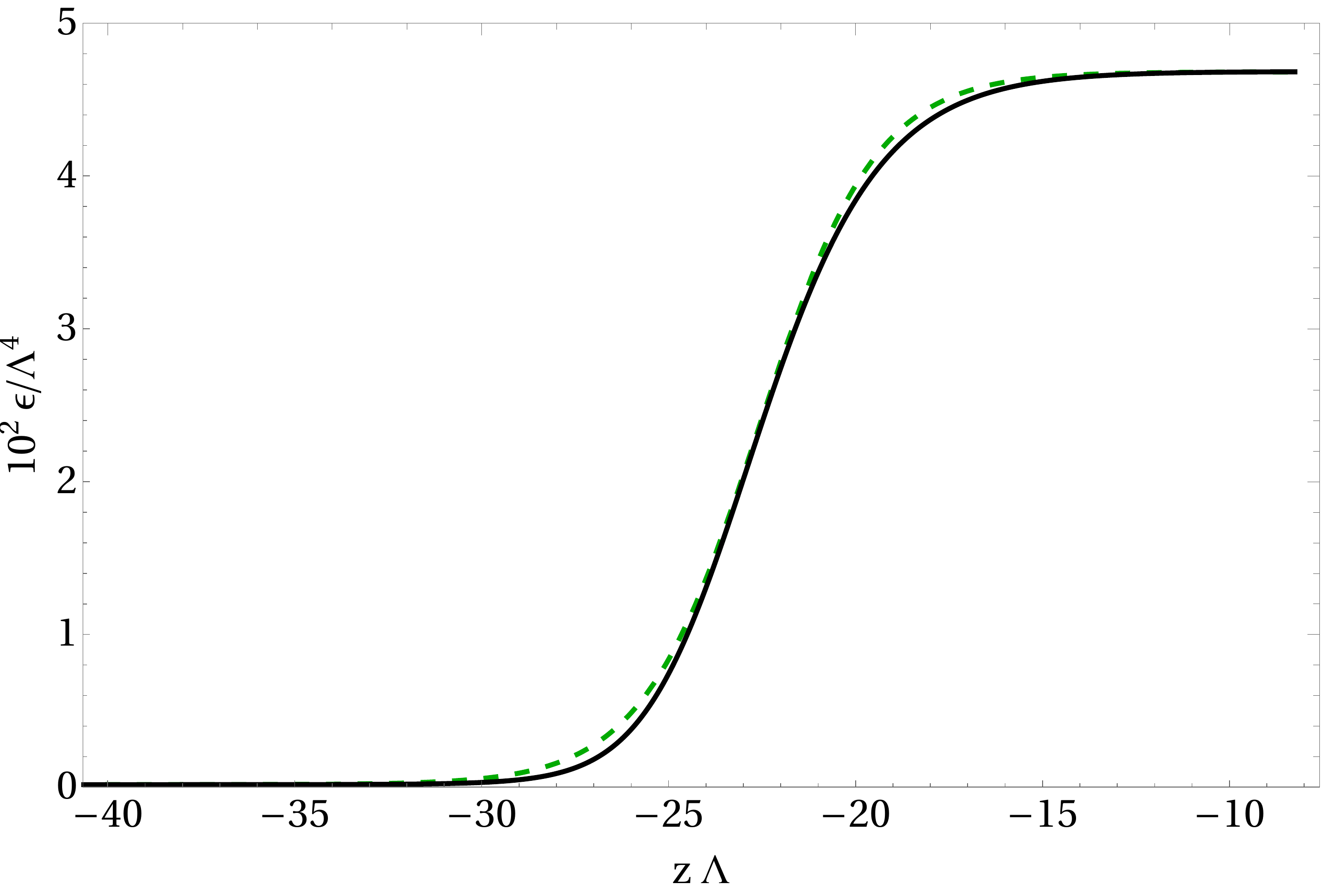}
\quad \hspace{-2mm}
\includegraphics[width=.51\textwidth]{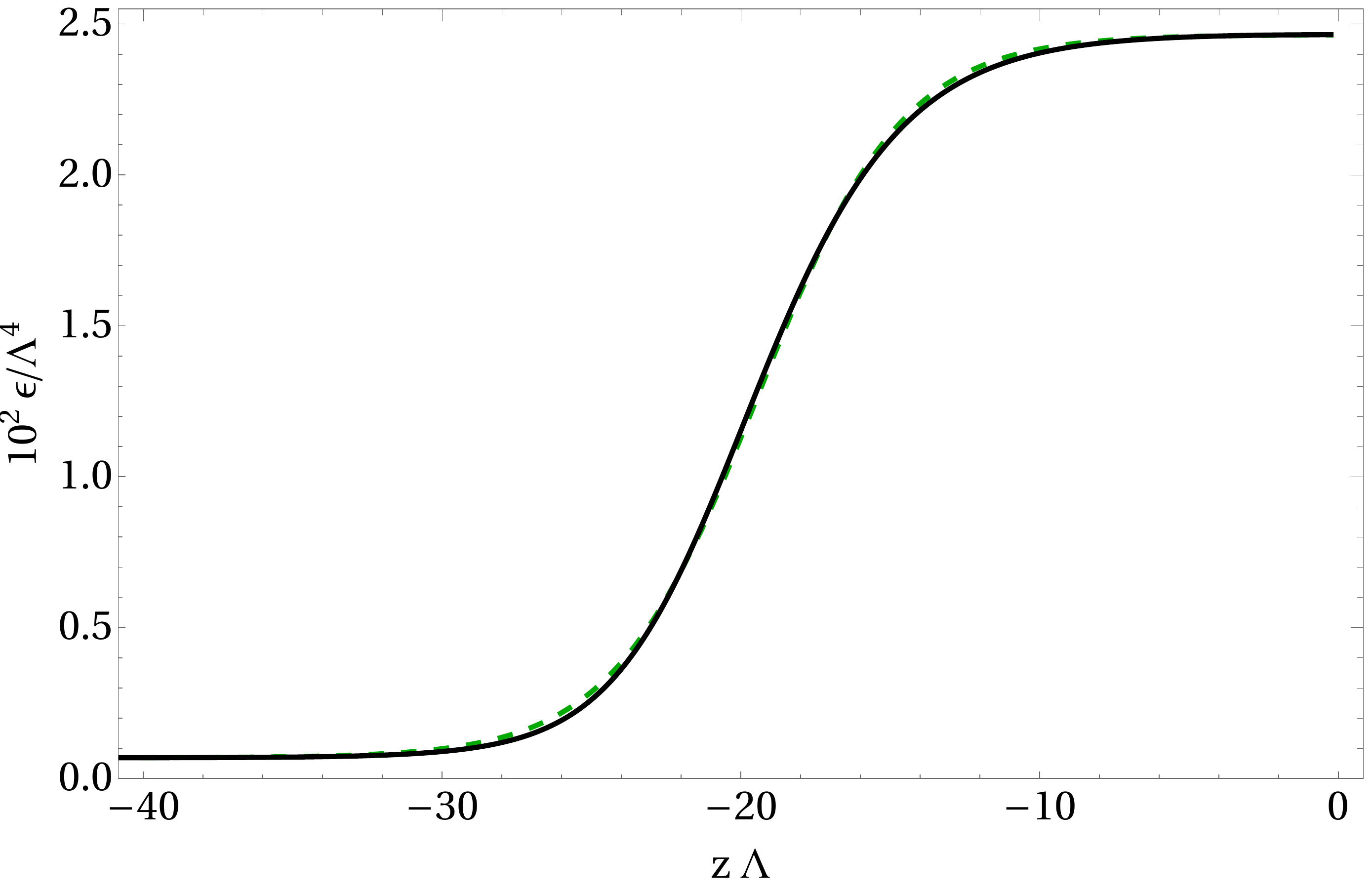}
\end{tabular}
\end{center}
\vspace{-5mm}
	\caption{\label{fig:domainshape} Longitudinal profiles of an interface with (left) $\phi_{\rm M} = 2.35$ and initial energy density $\E_3$; (right)  $\phi_{\rm M} = 2.45$ and initial energy density $\E_1$ both for a single plateau in continuous black with initial energy density $\E_3$  and the fitted interface shape in dashed green}
\end{figure*}
As seen in Fig.~\ref{fig:domainshape} the shape of the phase transition is well
approximated by the function
\begin{align}\label{eq:shape}
\E (z) \approx \frac{\Delta \E}{2} \left[ 1 + \tanh \left( \frac{z - z_0}{b} \right) \right] + \Elow
\end{align}
with $\Delta \E = \Ehigh - \Elow$, $z_0$ the exact midpoint between $\Elow$ and $\Ehigh$ of the interface and $b$ corresponds to the extent of the interface.
Note that the fit with Eq.~\ref{eq:shape} is very good for the interface from the midpoint $z_0$ to $\Ehigh$, but is only approximate on the downside approach from $z_0$ to $\Elow$.
The lower tail fit improves considerably for softer interfaces.
This improvement is visible in comparing Fig.~\ref{fig:domainshape} left and right plots.
For the interface of the domain wall with a confining vacuum~\cite{Aharony:2005bm} the $\tanh$ function is an almost perfect fit~\cite{jarvinen:20200520}.\\
\begin{table}[h!]
  \begin{center}
    \caption{Approximate width of the plateau interface.}
    \label{tab:width}
    \begin{tabular}{l|l|l|l|l|l}
   \toprule
$\boldsymbol{\phi_{\textrm M}}$ & $2.25$ & $2.3$ & $2.35$ & $2.4$ & $2.45$\\[0ex]
      \midrule
    $b \Lambda$ & $2.47$ & $2.75$ & $3.12$ & $3.62$ & $4.84$\\
   \bottomrule
    \end{tabular}
  \end{center}
\end{table}

As intuitively given by the picture of a less pronounced first-order phase transition the interface grows with more criticality.
There is almost a double increase of the extent going from $\phi_{\textrm M} =2.25$ to $\phi_{\textrm M} =2.45$ as listed in Table~\ref{tab:width}.
Counterintuitively, this means the spacetime of the simulated box with the same amount of total energy density needs to be bigger for the subcritical phase transition to fit nicely inside.
Nevertheless, it is hot stable energy density $\Ehigh$ in the studied theories,
which is the dominating thermodynamical quantity for determining how much energy
a phase separated final plateau has.
In all cases roughly more than $90\%$ of the total energy density concentrate
on the domain of the plateaux.
While less than $10\%$ of the total energy density are found in the tails
from the midpoint on.
Of course, this relates to the small cold energy density $\Elow$ and the
comparison assumes also the same finite longitudinal extent.
The steepness of the phase transition is the dominating factor leading
to only a small portion of the total energy density on the tails of the
interface.

\subsection{Surface tension}
With increasing criticality, the pressure exercised by the interface decreases.
This can be seen both by the decrease of the transverse pressure minima
at phase transition and directly by the surface tension of the interface:\\

The surface tension of the interface is per definition the excess free energy
per unit area in the transverse directions $x_\perp$.
The surface tension of the interface is positive.
For an equilibrated homogeneous system, the free energy density per unit volume is equal to the transverse pressure $F(z) = - P_T(z)$.
As discussed in \cite{Attems:2019yqn} by integrating the transverse
pressure minus the transition pressure over the full extent and taking
into account that the final solution has two interfaces, one gets the
surface tension of the respective interface
\begin{align}
\sigma
= \frac{1}{2} \int_0^{L_z} dz \left[ F(z) - F_c \right]
= -\frac{1}{2} \int_0^{L_z} \left[ P_T(z) - P_c \right] \,.
\end{align}
As expected, any presence of an interface increases the free energy of the system.\\
\begin{table}[h!]
  \begin{center}
    \caption{Surface tension in equilibrium and the absolute value
	  of the minima\\ of the transverse pressure for each interface.}
    \label{tab:surface}
    \begin{tabular}{r|r|r|r|r|r}
      \toprule
    $\boldsymbol{\phi_{\textrm M}}$ & $2.25$ & $2.3$ & $2.35$ & $2.4$ & $2.45$\\[0ex]
      \midrule
	    $\boldsymbol{\sigma /\Lambda^3}$ & $3.9 \times 10^{-3}$ & $2.9 \times 10^{-3}$ & $1.9 \times 10^{-3}$ & $1.1 \times 10^{-3}$ & $5.4 \times 10^{-4}$\\
      \midrule
    $|\boldsymbol{Min(P_T)|/\Lambda^4} $ & $7.9 \times 10^{-4}$ & $5.3 \times 1
	    0^{-4}$ & $3.1 \times 10^{-4}$ & $1.5 \times 10^{-4}$ & $3.4 \times 10^{-5}$\\

      \bottomrule
    \end{tabular}
  \end{center}
\end{table}
As listed in Table~\ref{tab:surface}, each theory gives rise to a distinct value.
The surface tension decreases by almost an order of magnitude from the
strongest to the softest first order phase transition.
Likewise the minima of the transverse pressure at each interface, as seen in
+able~\ref{tab:surface}, differ by more than an order of magnitude.
This indicates that the range of simulated theories vary widely in criticality.

\section{Evolution of the spinodal instability}
After the outline of the static properties of the inhomogeneous state
with varying criticality, I will demonstrate in what follows novel
dynamical evolution of the spinodal instability.

\subsection{Comparison of the formation time}
\begin{figure*}[t]
\begin{center}
\begin{tabular}{cc}
\includegraphics[width=.49\textwidth]{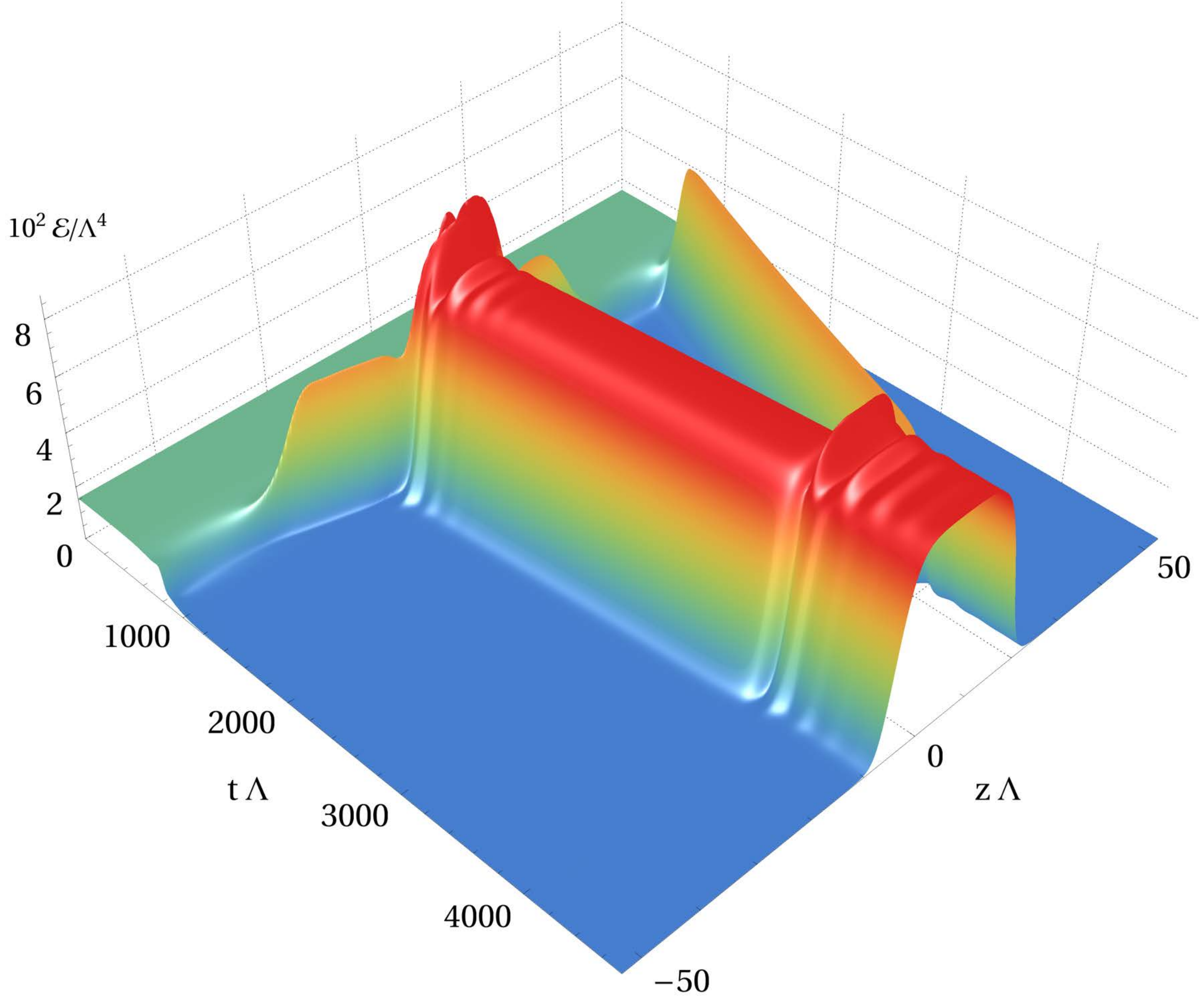}
&
\includegraphics[width=.49\textwidth]{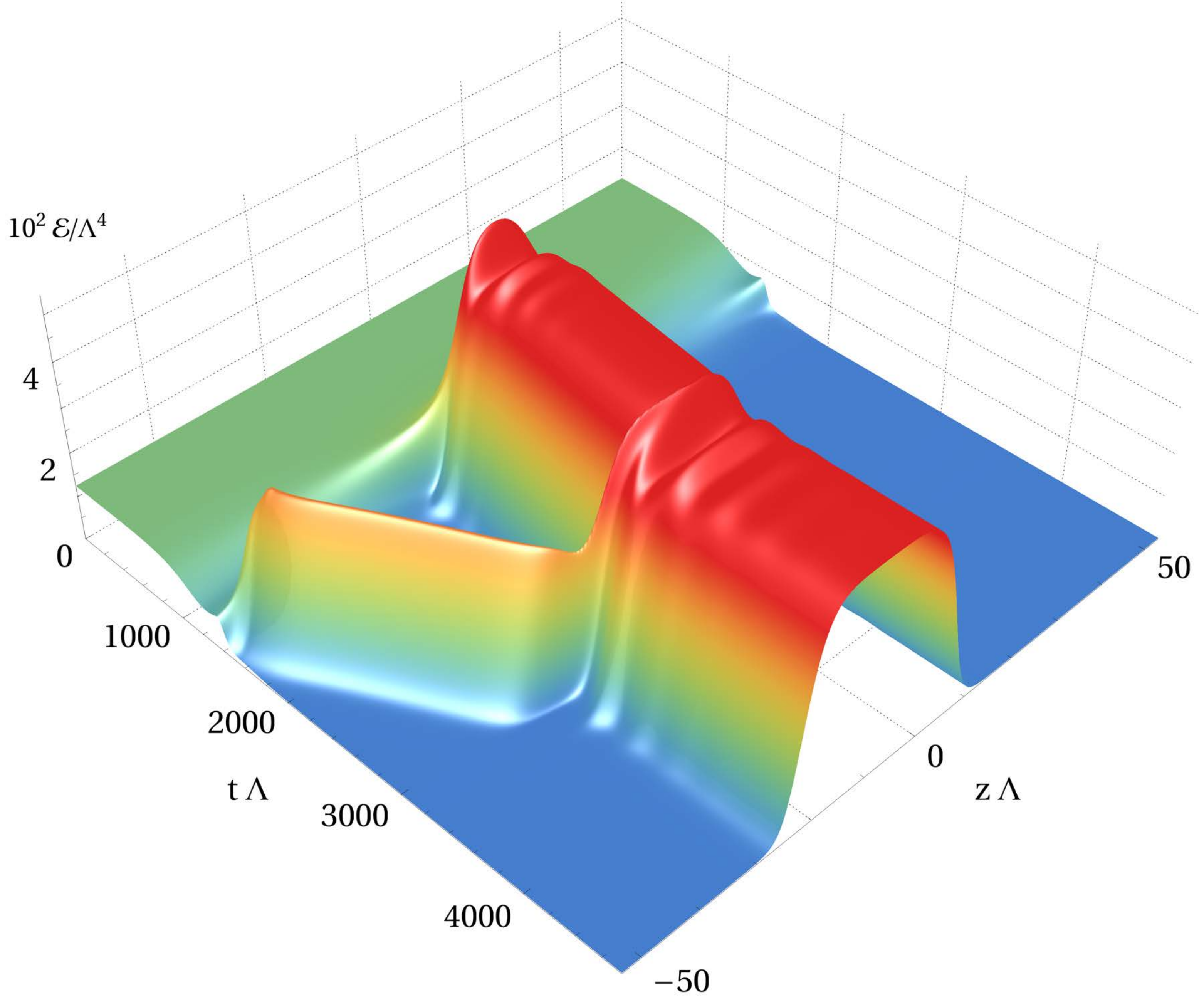}
\\[0.4cm]
\includegraphics[width=.49\textwidth]{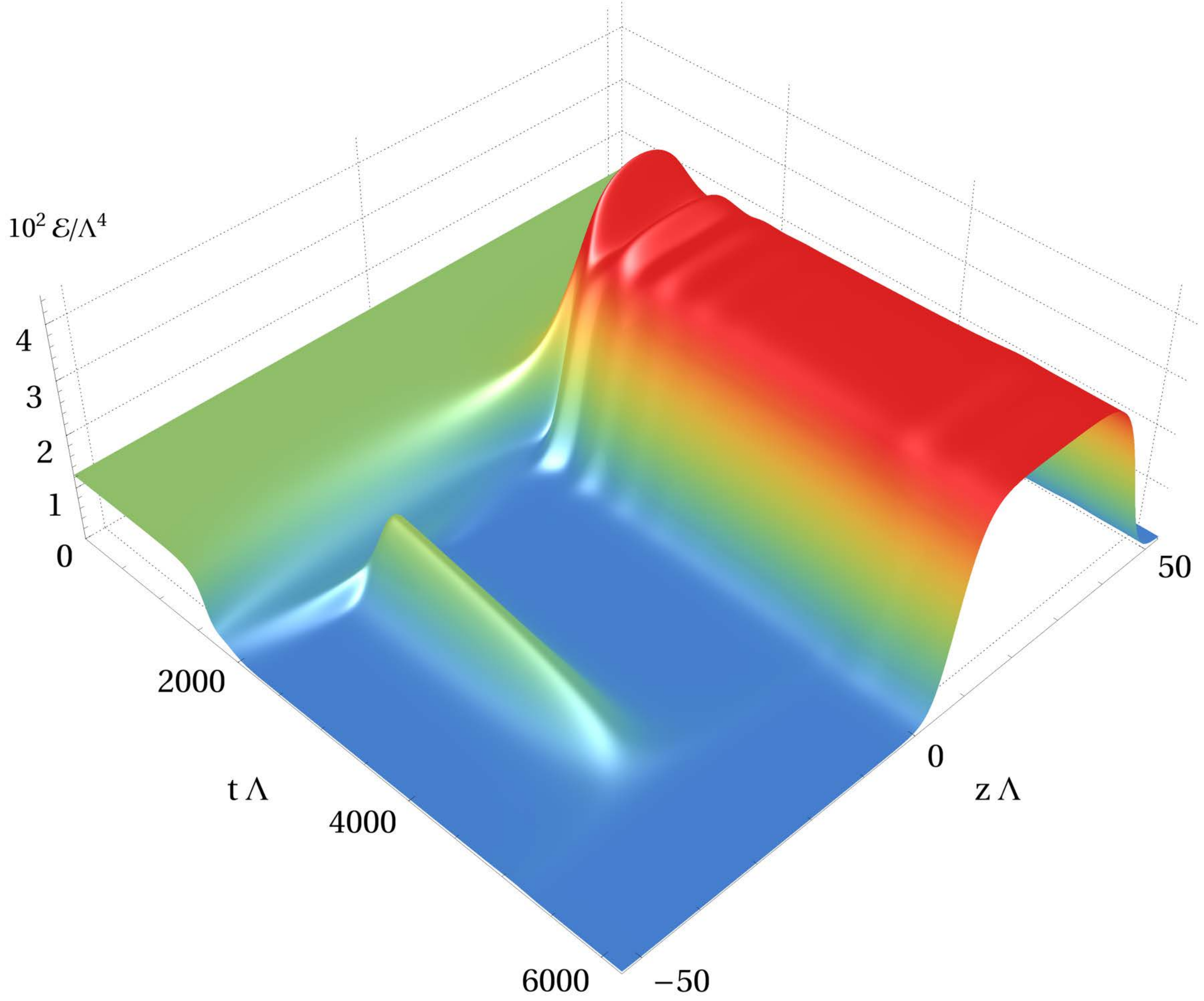}
&
\includegraphics[width=.49\textwidth]{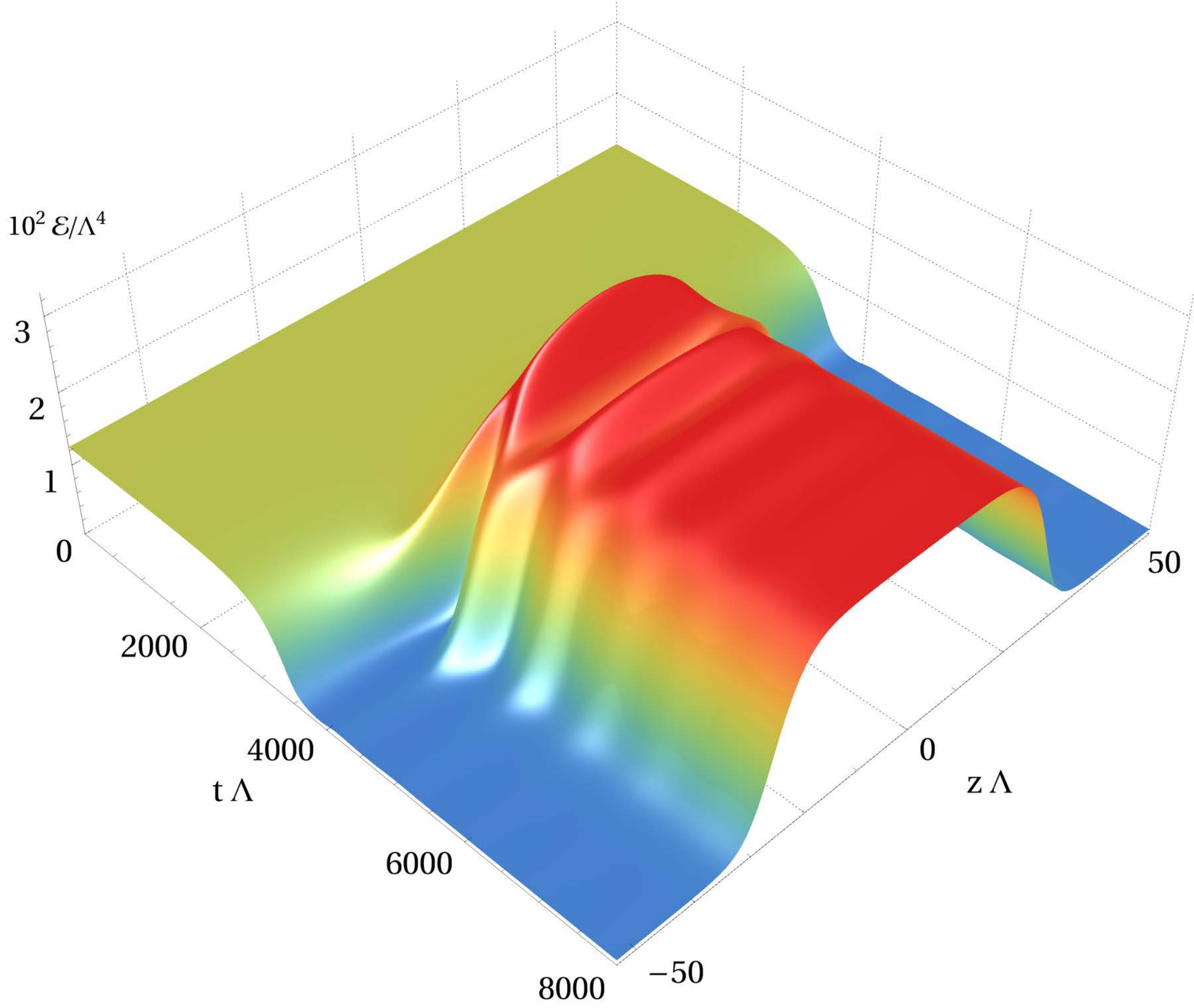}
\end{tabular}
\end{center}
\vspace{-5mm}
	\caption{\label{plateauformationquad} Four spacetime evolution of the local energy density with a plateau formation for the subcritical theories from top left to bottom right with $\phi_{\rm M} = \{2.25, 2.35, 2.4, 2.45\}$, initial local energy density $\E (t= 0) = \{\E_3, \E_2, \E_2, \E_2\}$ and same longitudinal extent $L_z \Lambda \approx 107$, where the plateau forms before merger with peaks (top left and right) or the plateau forms with a peak dissipation (bottom left) or only directly a plateau reshapes (bottom right).}
\end{figure*}
The more critical, meaning the closer the theory is near a critical point,
the slower the spinodal instability is.
This seems apparent from the different thermodynamics as
the unstable/spinodal region shrinks and the magnitude of the negative speed
of sound squared decreases too.
Summing it up: the more critical the theory is, the fewer unstable
modes it contains and the unstable modes have a slower growth rate.
The challenge here lies in unveiling this hypothesis in the full non-linear
situation of the spinodal dynamics and to show the differences with softer and
stronger phase transitions.\\

As a new measure of the dynamics of the spinodal instability I
consider the formation time, defined as the time from the start of the evolution
to when the system first reaches one of the two stable phases.
At formation time the inhomogeneous system reaches either $\Ehigh$ or $\Elow$.
As discussed in subsection \ref{subsec:reshaping}, this happens quite far from
equilibrium during the reshaping stage.
When the system has a large longitudinal extent this can be more complicated,
as the stable phase is only reached after the merger of several peaks
(seeing each of them is likely to be formed below $\Ehigh$).
Moreover, the formation time would depend on the initial
conditions and specifically on the initial velocity of the merging peaks.
them slower or faster merger.
The extraction of the formation time is approximate related to the fact,
that one can setup scenarios specifically triggering the most unstable mode
(instead of the first mode $n = 1$, which populates all unstable modes)
or initializing with stronger initial perturbation,
where the extracted formation time could be faster.

To extract the formation time and for the sake of comparing different theories,
one chooses a subset of simulations, where the spinodal instability directly
forms at least one plateau and has similar initial homogeneous local energy
density.
This setups allow for a direct comparisons of the formation time
of the spinodal instability with varying criticality.
Due to the simulations having the initial energy density below, but of the
similar magnitude of $\Ehigh$, usually the formation of a plateau with $\Ehigh$
happens long before the cool down of a valley to $\Elow$.
The time where the system reaches both $\Elow$ and $\Ehigh$ is the full phase
separation time.
This full separation time depends heavily on the initial state and thus needs
a much higher number of different simulations for a meaningful comparison.
As both the cooled and the heated regions overshoot the respective
equilibrium values of $\Elow$ and $\Ehigh$ no tolerance criterium is applied
for the computation of the formation time.
In what follows I extract the direct formation time of a plateau.

\begin{table}[h!]
  \begin{center}
  \caption{Formation time of the spinodal instability with varying criticality}
    \label{tab:time}
    \begin{tabular}{l|l|l|l|l}
      \toprule
	    $\boldsymbol{\phi_{\textrm M}}$ & $2.25$ & $2.35$ & $2.4$ &$2.45$ \\[0ex]
      \midrule
	    $\boldsymbol{t_{\textrm formation} \Lambda}$ & $880$& $1380$ & $1725$& $2660$ \\
      \bottomrule
    \end{tabular}
  \end{center}
\end{table}
The runs visualized in Fig.~\ref{plateauformationquad} provide from all the
available simulations the direct plateau formation for each theory
with similiar initial state.
The formation time of the spinodal instability in the simulated theories triples near criticality as documented in the Table~\ref{tab:time}.
In all the simulations of the Table~\ref{tab:time} the spinodal instability is triggered by the same small initial perturbation.
Each simulation has also enough total energy density to form a plateau.
Of course the theory with the strongest phase transition with $\phi_{\rm M} =2.25$ has the fastest formation time.
The softest theory near criticality with $\phi_{\rm M} =2.45$ has a
triple longer formation time.

For the theory with $\phi_{\rm M} = 2.3$, no clean direct plateau formation has
been simulated, hence it is not listed in the formation Table~\ref{tab:time}.
Nevertheless its fastest extracted formation time of $t_{\textrm formation}
\Lambda = 1190$ fits in the overall trend of the table.
In the plateau merger of Fig.~\ref{mergers} one extracts the fastest respective formation time of $t_{\textrm formation} \Lambda = \{ 1460, 1420 \}$ for $\phi_{\rm M} =2.3$.
Due to forming multiple peaks and having to wait for first mergers these
formation times are slower and not representative for the corresponding theory.
In Fig.~\ref{dissipation}(left) with $\phi_{\rm M} = 2.4$,
one gets roughly a formation time of $t_{\textrm formation} \Lambda = 2400$
and in Fig.~\ref{dissipation}(right) with $\phi_{\rm M} = 2.35$ also of
$t_{\textrm formation} \Lambda = 2400$.
Here in both cases the slowdown, relative to the extracted values in Table~\ref{tab:time} happen due to a double formation of a
plateau and a peak.
In conclusion for a meaningful comparison of the formation time, one needs
to simulate direct single plateau formations.
Aptly the strongest phase transition with the largest spinodal region
shows the fastest formation time.

\subsection{Peak dissipation into plateaux}
An open puzzle of the holographic simulations of the spinodal instability
is the rigidity of the interface.
While peaks and plateaux happen to move in the longitudinal extent,
their interfaces keep a fixed shape.
Even in collisions, the interface of the phase separation only wobbles in a very
rigid way.
Moreover, only over an extended spacetime have the peaks a slowly varying velocity with almost no distortion of their shape.
Here are presented the first calculations of a peak dissipating into a plateau
for two different theories.
This shows the possibility of inhomogeneous states to change their shape.\\
\begin{figure*}[t]
\begin{center}
\begin{tabular}{cc}
\quad \hspace{-2mm}
\includegraphics[width=.49\textwidth]{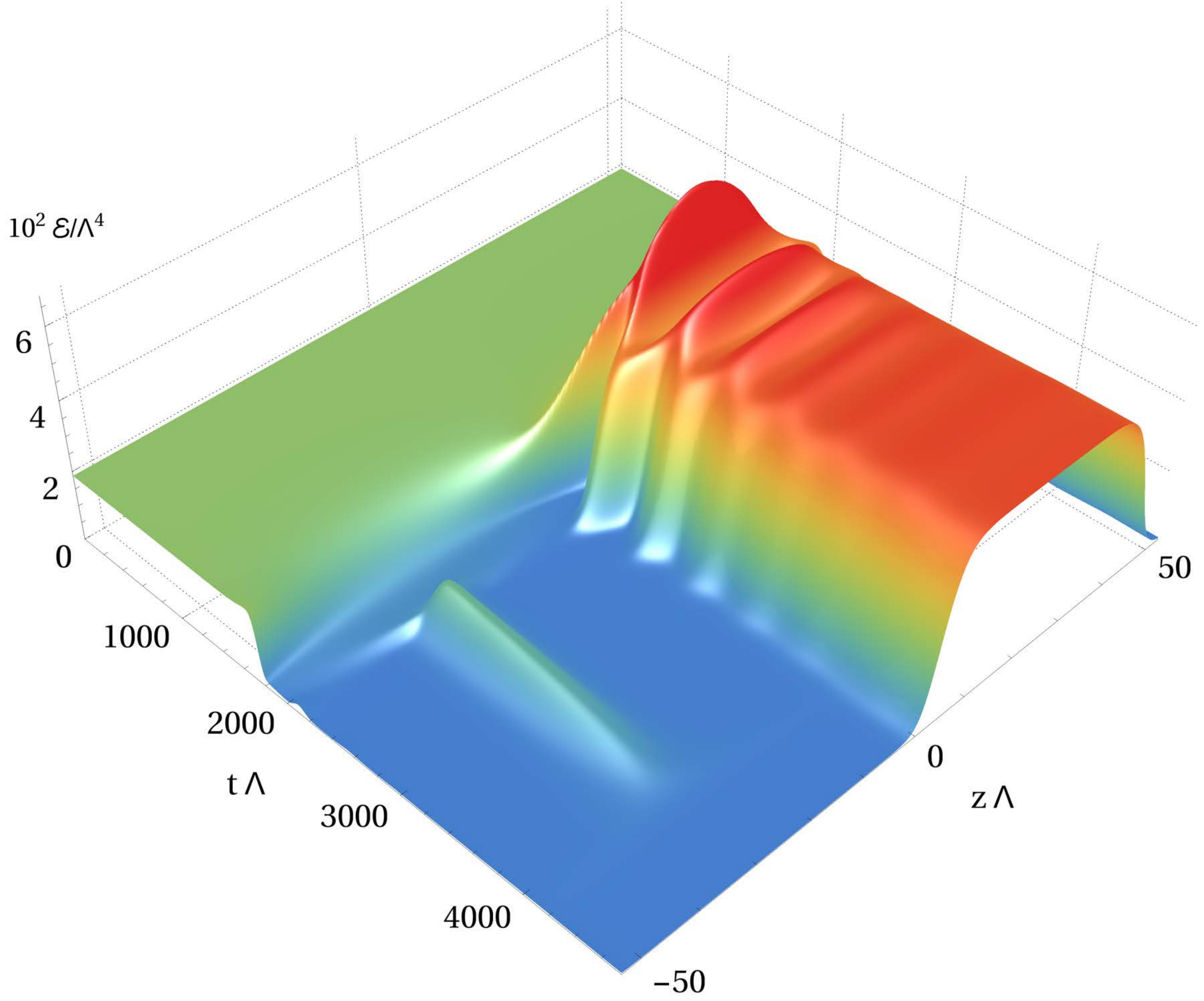}
&
\includegraphics[width=.49\textwidth]{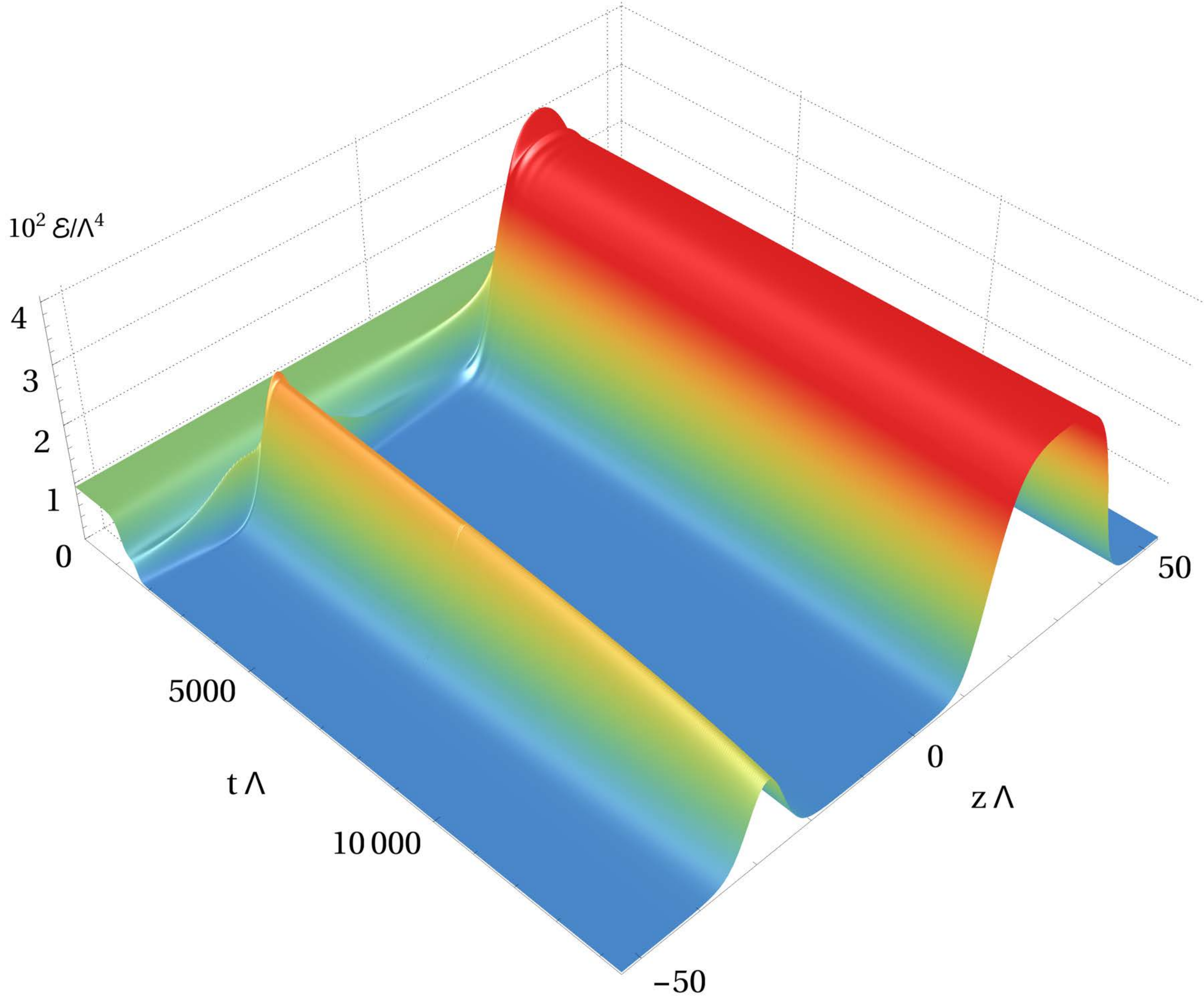}
\end{tabular}
\end{center}
\vspace{-5mm}
	\caption{\label{dissipation} Two subcritical spacetime evolution of the
	local energy density demonstrating a dissipation of a peak into a plateau
	with $L_z\lambda = 107$,
	(left) initial homogeneous $\E(t=0)= \E_4$ and $\phi_{\rm M} = 2.35$ up to $t \Lambda = 6250$;
	(right) initial homogeneous $\E(t=0)= \E_1$ and $\phi_{\rm M} = 2.4$ up to $t \Lambda = 14500$.}
\end{figure*}

This process has been newly observed and needs a finely tuned setup.
The system needs to directly produce a plateau and a peak both with no
initial velocities.
In Fig.~\ref{plateauformationquad}(bottom left) and in Fig.~\ref{dissipation}(left $+$ right) are shown three systems for different theories
$\phi_{\rm M} = \left\{2.35, 2.4\right\}$ and initial energies, but each with
enough total energy density to each form a plateau and a peak.
A similar fine-tuned setup can also be reproduced in a theory with a very strong
first order phase transition with $\phi_{\rm M} = 2.3$.
Therefore this untypical dissipation process does not depend on criticality.
In the number of simulations\cite{attems_maximilian_2020_3445360,attems_maximilian_2019_3445360} performed up to date this
process has not yet been observed in large extents, where several peaks or
plateaux initially form.
While peaks with a slow velocity form on a regular basis it is quite untypical
for the spinodal instability to morph into quasi-static features.
This happens only if an excited dominant unstable mode carries over from the
first stage into the reshaping stage forming a certain number of quasi-static
peaks.
Then after some little perturbation these peaks usually gain velocity to merge.
In this tuned setup, the peak diminishes while staying at the same location.

In Fig.~\ref{plateauformationquad}(bottom left) the simulation first forms both a plateau at around $t \Lambda \approx 1725$
and a peak around $t \Lambda \approx 2250$.
Surprisingly, the peak stays at the same spacetime with zero velocity and its
maximum energy density decreases first quasi-linear over $t \Lambda \approx
2250$ time duration and then exponentially for $t\Lambda \approx 600$ until it
completely disappears.
During the quasi-linear stage of the dissipation of the peak, the plateau grows
in total energy density, while the valley between the peak and plateau
continues to exist.
The valley fills up during the exponential stage of the dissipation
precipitating the decay of the peak.
Afterwards the minimum again cools down to the cold stable phase.
After $t \Lambda \approx 5100$ the last amount of local energy density joins
and slightly perturbes the plateau.

In Fig.~\ref{dissipation}(left) one sees first the formation of a plateau
at around $t \Lambda \approx 1900$ and a stationary peak around $t \Lambda \approx 2200$.
At first the dissipation happens in quasi-linear form during $t\Lambda \approx 1300$
and then exponentially over $t\Lambda \approx 700$.
The first stage is quasi-linear as on top of the dissipation proceeds the
fast equilibration of the peak after its formation.
This dissipation process takes less time as the dissipation in Fig.~\ref{plateauformationquad}(bottom left),
as this peak initially has approximately a quarter less total energy density
and the quasi-linear dissipation is faster.
Again the end of full dissipation of the peak around $t \Lambda \approx 4200$ one
notices a slight perturbation of the plateau as the last energy density joins.
Here this perturbation is more difficult to discern in the visualization as
the plateau itself is still equilibrating from its formation.

In Fig.~\ref{dissipation}(right) one notices the formation of a plateau at
around $t \Lambda \approx 1500$ and then of a stationary peak at around
$t \Lambda \approx 2050$.
While this would be the fastest formation time for $\phi_{\rm M} = 2.4$,
the available dataset has no comparable simulations for any other theories
with similar initial state, hence it was not considered in Table~\ref{tab:time}.
The third case has the slowest dissipation with the peak only loosing one
third of its height over the duration of more than $t \Lambda \approx 10000$.
The exponential dissipative stage sets in at around $t \Lambda \approx 12500$.
In this simulation the dissipation is still ongoing at $t \Lambda =14500$ and
will take more computing time to fully dissipate.\\

All three presented atypical cases have a plateau forming before a stationary
peak, which dissipates first in linear and than exponentially manner.
In conclusion this atypical setup happens for a system with enough total
energy density to form a plateau and a peak, but not another inhomogeneous
structure.
It is striking that the dissipation happens independently of criticality both
for harder and softer phase transitions.
It is the only process known that defies the rigidity of the interface.

\section{Discussion}
Simulating the spinodal instability far from and near a critical point,
one recognizes that the harder the first order phase transition is the more
dramatic the spinodal instability is.
The shrinking extent and the harder surface tension of the interface of
the phase separation go pair in pair.
New insights from the dynamical simulations are the rising formation time,
while showcasing similar merger dynamics with varying degree of criticality.\\

Looking at the holographic dynamics of the spinodal instability one remarks:
The more pronounced the first-order phase transition, the more unstable modes
there are which dictate a faster spinodal instability.
This translates in a much faster formation time away from criticality.
With decreasing criticality the domain wall between the stable phases shrinks.
As the magnitudes between the local energy densities $\Elow$ and $\Ehigh$ grows
with less criticality, one needs more local energy density in the system to
generate a phase separated solution.
The finding of this single preferred final solution\cite{Attems:2019yqn} is now
substantiated with softer and harder first-order phase transitions.
The newly proposed criterium for the distinction of this inhomogeneous state
preforms well in all stages of the spinodal instability.
It fittingly distinguishes between the phase separated plateaux and prevalent
local maxima, so named peaks.\\

Moreover the newly found atypical setup with a peak dissipating into a plateaux
is a progress in the understanding of the rigidity of the interface.
For the first time, an inhomogeneous solution has been shown to change its
shape after formation without directly merging.
The peak without velocity slowly dissipates.
This setup shows that the interface is not fully immovable.
It is still surprising how rigid the interface is across the theories with
widely different criticality.
An open question remains to simulate a collision with high velocities,
where the remnants would maybe separate again post collision.
For now, all scenarios of the spinodal instability demonstrate peaks joining
together with no side remnants.
A possible scenario involves a peak with high speed crossing a plateau.
It would also be interesting to be able to trigger the spinodal instability in
a plasma blob created by shockwave collisions near a critical point.
The numerical challenge here lies in the extreme slow down of the dynamics
at a critical point~\cite{Attems:2018gou}.\\

Although this approach to criticality used a bottom-up model, the performed
simulations suggest that the qualitative physics may be quite universal
in strong coupling situations approaching a critical point where the negative
speed of sound induces the spinodal instability.
It would be interesting to break the planar symmetry to allow dynamics
in the transversal directions.
While the reshaping stage dynamics may change, the simulated formation times
will be similar to this study.
Extending the simulations to include conserved charges such as the chemical baryonic
potential $\mu$ would be interesting to enhance the analysis to the full
$T\mu$ phase diagram~\cite{Critelli:2017oub,Du:2020bxp,Dore:2020jye} and
make contact with the experimental programs:
A major goal for heavy-ion collisions is to determine the existence
of a first-order phase transition between between hadronic and quark-gluon
plasma in the QCD phase diagram~\cite{Fukushima:2010bq,Fukushima:2013rx}, which is predicted
by numerous effective field theory models~\cite{Stephanov:2004wx}.
As the first order transition presumably has a big temperature range it is worthwhile
to explore if the heavy-ion experiments see a signature~\cite{Bluhm:2020mpc} of it.
Very recently a sophisticated hadronic mean-field simulation using the Vlasov
equation has also shown that hadronic systems initialized in unstable regions
of the phase diagram undergo spontaneous spinodal
decomposition~\cite{Sorensen:2020ygf}.
It will be exciting to see the upcoming comparisons with experimental data
influenced by the two phase system.\\

Previously we have demonstrated that the spinodal instability is very well
described by a second-order hydrodynamics~\cite{Attems:2019yqn} with second-order purely spatial
gradients included and redefined non-conformal second-order transport coefficients.
It would be very interesting to develop the new hydrodynamics formulation
of~\cite{Bemfica:2017wps,Kovtun:2019hdm}, which may be able to incorporate
the large spatial gradients of the phase separation for a full hydrodynamical
evolution.

\section*{Acknowledgments}\label{sec:ack}
I am grateful for fruitful discussions with J.~Casalderrey-Solana, E.~Kiritsis and D.~Mateos;
Y.~Bea, J.~Brewer, N.~Jokela, A.~Vuorinen and M.~Zilhao for interesting exchanges and
R.~Janik, M.~Jarvinen, M.~Stephanov and L.~Yin for general discussions on the critical point and phase transitions.
I thank the Center for Supercomputing of Galicia (CESGA) for providing extensive
High Performance Computing resources in the Finisterra cluster (project usciemat).
I would like to thank the Institute for Nuclear Theory for their
hospitality at the INT-19-1b workshop during early stages of this work.
I acknowledge support through H2020-MSCA-IF-2019 ExHolo 898223.
This work is also supported by the “María de Maeztu” Units of Excellence program MDM-2016-0692, Xunta de Galicia (Consellería de Educación) and the Spanish Research State Agency.

\bibliographystyle{utphys}
\bibliography{ref}

\end{document}